\documentclass[twocolumn]{aastex631}

\usepackage{amsmath}
\usepackage[version=4]{mhchem}
\usepackage{verbatim}
\usepackage{bm}
\usepackage{soul}

\shorttitle{CIDA 9A: Dead or Alive?}

\begin{document}

\title{Dual-Band Observations of the Asymmetric Ring around CIDA 9A: Dead or Alive?}

\author[0000-0001-6307-4195]{Daniel Harsono}
\affiliation{Institute of Astronomy, Department of Physics, National Tsing Hua University, Hsinchu, Taiwan}

\author[0000-0002-7607-719X]{Feng Long}
\altaffiliation{NASA Hubble Fellowship Program Sagan Fellow}
\affiliation{Lunar and Planetary Laboratory, University of Arizona, Tucson, AZ 85721, USA}
\affiliation{Center for Astrophysics \textbar\ Harvard \& Smithsonian, 60 Garden St., Cambridge, MA 02138, USA}

\author{Paola Pinilla}
\affiliation{Mullard Space Science Laboratory, University College London,
Holmbury St Mary, Dorking, Surrey RH5 6NT, UK}

\author{Alessia A. Rota}
\affiliation{Leiden Observatory, Leiden University, P.O. Box 9513, NL-2300 RA, Leiden, The Netherlands}

\author{Carlo F. Manara}
\affiliation{1European Southern Observatory, Karl-Schwarzschild-Strasse 2, D-85748 Garching bei München, Germany}

\author{Gregory J. Herczeg}
\affiliation{Kavli Institute for Astronomy and Astrophysics, Peking University, Yiheyuan 5, Haidian Qu, 100871 Beijing, China}
\affiliation{Department of Astronomy, Peking University, Yiheyuan 5, Haidian
Qu, 100871 Beijing, China}

\author[0000-0002-6773-459X]{Doug Johnstone}
\affiliation{NRC Herzberg Astronomy and Astrophysics, 5071 West Saanich Road, Victoria, BC, V9E 2E7, Canada}
\affiliation{Department of Physics and Astronomy, University of Victoria, 3800 Finnerty Road, Elliot Building, Victoria, BC, V8P 5C2, Canada}

\author{Giovanni Rosotti}
\affiliation{School of Physics and Astronomy, University of Leicester, Leicester LE1 7RH, UK}
\affiliation{Leiden Observatory, Leiden University, P.O. Box 9513, NL-2300 RA, Leiden, The Netherlands}

\author{Giuseppe Lodato}
\affiliation{Dipartimento di Fisica, Universitá degli Studi di Milano, Via Giovanni Celoria, 16, I-20133 Milano, MI, Italia}

\author{Francois Menard}
\affiliation{Univ. Grenoble Alpes, CNRS, IPAG / UMR 5274, F-38000 Grenoble}
checked and all good, on 1/3/23 by FMe

\author{Marco Tazzari}
\affiliation{Institute of Astronomy, University of Cambridge, Madingley Road, CB3 0HA Cambridge, UK}

\author{Yangfan Shi}
\affiliation{Department of Astronomy, Peking University, Yiheyuan 5, Haidian
Qu, 100871 Beijing, China}

\begin{abstract}
While the most exciting explanation of the observed dust asymmetries in protoplanetary disks is the presence of protoplanets, other mechanisms can also form the dust features.  This paper presents dual-wavelength Atacama Large Millimeter/submillimeter Array (ALMA) observations of a large asymmetric dusty ring around the M-type star CIDA 9A. We detect a dust asymmetry in both 1.3 mm and 3.1 mm data.  To characterize the asymmetric structure, a parametric model is used to fit the observed visibilities. We report a tentative azimuthal shift of the dust emission peaks between the observations at the two wavelengths.  This shift is consistent with a dust trap caused by a vortex, which may be formed by an embedded protoplanet or other hydrodynamical instabilities, such as a dead zone.  Deep high-spatial observations of dust and molecular gas are needed to constrain the mechanisms that formed the observed millimeter cavity and dust asymmetry in the protoplanetary disk around CIDA 9A. 
\end{abstract}


\section{Introduction} \label{sec:intro}

The Atacama Large Millimeter Array (ALMA) has uncovered the early stages of planet formation, with dust continuum observations that reveal rings, gaps, spiral arms, and asymmetric dust emission \citep[e.g.,][]{hltau2014, flong18, dsharp, huang18, cazzoletti18, cieza21}.  These features may be signatures of embedded planets or hydrodynamical instabilities may create them (see review by \citealt{baeppvii}) that are creating planetesimals.  Prime examples of these cradles of planet formation have been transition (or cold) disks \citep{espaillat14}, disks with a deep millimeter central cavity that are typically attributed to the presence of young giant planets \citep[e.g.,][]{duffell15,fung16,calcino20}. However, young giant planets have only been reported in three protoplanetary disks, PDS 70, AB Aur, and HD 169142 \citep{Keppler2018, muller18,currie22,hammond23}; deep observations of molecular gas show gas gaps and kinematics that may point to indirect evidence of protoplanets \citep{vandermarel16a, rdong17, fedele21,stadler23}.

Large millimeter dust cavities, rings, and large-scale azimuthal dust asymmetries are attributed to dust trapping induced by pressure bumps \citep[e.g.,][]{klahr97, pinilla12a}.  Dust particles marginally coupled to the gas tend to accumulate at the local pressure maxima of a protoplanetary disk \citep{weidenschilling77, birnstiel10}.  The growth of dust particles is enhanced inside the pressure trap leading to potential planetesimal formation \citep[e.g.,][]{youdin05,johansen07,drazkowskappvii}.   While many mechanisms can generate the pressure bumps \citep{baeppvii}, they are often associated with young Jovian planets, regardless of whether or not the planets are directly observable  \citep[e.g.,][]{ataiee13, espaillat14, rdong15b, logan20a}.  The interaction between the young planet and the disk can launch a spiral arm that is detectable in the infrared through scattered light \citep[e.g.,][]{bae16, rdong17, benistyppvii, paardekooperppvii}.  In such a scenario, the young planet is expected to carve a deep gap in the protoplanetary disk, leading to different gas and dust distributions where a substantial amount of gas is removed in some cases.  Hydrodynamical simulations show that the exact depletion factor depends on the number of planets, their orbital orientation, and the viscosity in the disk \citep[e.g.,][Bae et al. 2022]{duffell15, pinilla22}.

Alternative scenarios for the formation of large inner cavities are: a close binary \citep{artymowicz94}, photoevaporation \citep{matsuyama03, alexander07}, magneto-hydrodynamic winds \citep{tksuzuki10}, dead zones \citep{pinilla16, delage2022}, or a combination of these mechanisms \citep{rosotti13, garate21}. Each proposed mechanism has observable features that can be tested against spatially resolved dust and gas observations, however, there are degeneracies. For example, all of them are expected to trigger dust trapping at the outer edge of the cavity. Nonetheless, there are some distinct predictions for each of these mechanisms. For instance, an inner cavity formed by a dead zone may show a \emph{radial} shift between different sizes of dust grains that is dissimilar from the planet-disk interaction models. In the dead zone case, the location of the radial peak is closer at longer wavelengths (tracing larger grains) than at short wavelengths (tracing smaller grains) \citep[Figure 5 in][]{pinilla2019}. This difference can be tested by observing dust emission at two frequencies within the ALMA Bands.  On the other hand, a vortex generated by planet(s) inside the cavity will create an observable azimuthal shift in the dust millimeter emission \citep{Baruteau2016}, while it is still unclear if that is the case in the dead zone case. On the other hand, disk winds can be detected through broad ($>5$ km/s) molecular and atomic lines \citep[e.g.,][]{pascucci11,klaassen13, klaassen16, booth21}. Therefore, with high-spatial resolution observations of gas and dust emission at different wavelengths, it is possible to start ruling out one or more of the proposed mechanisms toward specific systems.

Thus far, a few disks show a clear azimuthal asymmetry in the millimeter dust emission.  Protoplanetary disks around IRS48 \citep{vandermarel13sci} and HD142527 \citep{casassus13nat} are iconic examples of azimuthally asymmetric dust disks revealed by ALMA.  Other disks show various dust sub-structures such as the dust ring morphology \citep{kraus17, boehler18, yliu19} or large horseshoe-like shape \citep{tang12, cazzoletti18, muroarena20,vandermarel21}.  By spatially resolving the disk around HD135344B, \citet{cazzoletti18} reported an azimuthal shift between observations at different wavelengths toward its dust arc, indicating a vortex.   Recently, \citet{boehler21} also demonstrated how deep observations of molecular lines can be used to study the kinematics around a large vortex in HD142527.

Most previous detections and characterizations of large azimuthal asymmetries have been found toward protoplanetary disks around Herbig stars.   In the Taurus disk survey by \citet{flong18}, the disk around the M2 primary star CIDA 9A shows a $\sim$ 25 au inner cavity and an asymmetric dust ring.  CIDA 9 (BCG93 9, \citealt{bcg93}) is a binary system with a separation of $2\farcs{35}$ \citep{manara19}, which corresponds to a projected separation of 411 au at a distance of 175 pc \citep[Gaia DR3,][]{gaiaedr3}. The central star has a lower mass than the other systems whose disks show a large millimeter cavity, providing a test for disk evolution around lower mass stars where dust radial drift is expected to be more efficient.  Although still not well studied, the general properties of disks around very low-mass stars tend to follow their solar-mass star counterparts \citep{kurtovic21}.  For example, the detection of a 20 au cavity around the M 4.5 star CIDA 1 suggests that a Saturn-mass planet is needed to open the gap \citep{pinilla18}.

We obtained ALMA dual-band observations of the disk around CIDA 9A to better understand the origin of its large millimeter cavity and asymmetric dust disk.  In this paper, we report on the detection of a large-scale azimuthal asymmetric ring in both Bands 3 (3.1 mm) and 6 (1.3 mm) around the primary star CIDA 9A, with an azimuthal shift in the asymmetry between the two wavelengths.  This paper is organized as follows. Section~\ref{sec:observations} describes the observational details and the synthesis imaging. The dust continuum and gas observations analysis can be found in Section~\ref{sec:results}. We discuss the implications of the observed substructures in the context of a vortex and its lifetime in Section~\ref{sec:discuss}. Finally, Section 5 summarizes the paper and the conclusions. 


\section{Observational details} \label{sec:observations}

\subsection{CIDA 9A stellar properties: mass and age}

The stellar mass and age estimates of CIDA9A are obtained by comparing the stellar isochrone to the stellar luminosity and temperature.  Based on fitting photosphere, accretion, and extinction to low-resolution optical spectra, CIDA 9A has a spectral type of M1.8 ($T_{\rm eff}=3592$ K and luminosity of $0.21$ $L_\odot$, \citealt{herczeg14}), after accounting for an extinction $A_V=1.35$, contribution from accretion, and corrected for $d=175$ pc \citep{gaiaedr3}.  The automated temperature measured from LAMOST spectra of $3675$ K is slightly higher \citep{luo19}, perhaps because the fit to the LAMOST spectra did not include an accretion component.  A similar luminosity is retrieved when using PAN-STARSS $y$ and $z$-band photometry \citep{chambers16} and color and bolometric corrections by \citet{pecaut13} and \citet{zhou22}.

These parameters lead to a mass of $\sim 0.42$ M$_\odot$ and age of $3.5$ Myr, when compared to standard evolutionary models (\citealt{baraffe15} and \citealt{somers20} without spots), and $\sim 0.66$ M$_\odot$ and age of $\sim 6$ Myr for models that include non-standard physics (the magnetic models of \citealt{feiden16} or models from \citealt{somers20} with 50\% spot coverage).  If the stellar mass is $\gtrsim 0.7$ M$_\odot$, the stellar age would be as high as 6--10 Myr.  However, the radius may be underestimated if the star is heavily spotted, which can lead to an overestimation of the effective temperature for the visible surface \citep{gully17}.  The brightness of CIDA 9A varies from $V=15$ to 17 mag in ASAS-SN monitoring \citep{shappee14,kochanek17}, which also introduces significant uncertainties in any radius estimate.

 While most of the Taurus young stellar objects are at a distance of 130--200 pc \citep{pabgalli19, krolikowski19} and ages between 1 -- 3 Myr, there are some apparent outliers and older subclusters with ages up to $\sim$20 Myr \citep{kraus17,krolikowski19,liu21}.  It seems plausible that CIDA 9 is an old system in the Taurus Molecular Cloud, although age estimates are uncertain on any individual star \citep[see uncertainties described by][]{soderblom14}, especially one with such high amplitude variability.

The accretion luminosity of 0.026 $L_\odot$ is measured from the excess hydrogen continuum emission in a flux-calibrated optical spectrum that covers the Balmer Jump (from the spectrum described in \citet{herczeg14} and following the methods described in \citealt{herczeg08}).  This accretion luminosity corresponds to an accretion rate of $1.8\times10^{-9}$ M$_\odot$ yr$^{-1}$, for the adopted parameters 1.18 R$_\odot$ and 0.66 M$_\odot$.


\begin{table}[]
    \centering
    \caption{Targeted spectral windows in Band 3 and molecular lines.}
    \label{tab:spws}
    \begin{tabular}{c c c c c}
    \hline
    \hline
         SPW & $\nu_0$ & $d\nu$ & $dv$ & Transitions  \\
            &  (GHz) & (kHz) & (km s$^{-1}$)& \\
    \hline 
    1   & 97.976   & 61.035  & 0.19 & \ce{CS 2-1} \\
    2   & 97.695   & 61.035  & 0.19 & ... \\
    3   & 95.999   & 976.562  & 3.0 & ... \\
    4   & 109.777   & 61.035  & 0.17 & \ce{C^{18}O 1-0} \\
    5   & 110.196   & 61.035  & 0.17 & \ce{^{13}CO 1-0} \\
    6   & 107.894   & 976.562  & 2.7 & ... \\
    \hline
    \end{tabular}
\end{table}

\subsection{ALMA observations}

ALMA observed CIDA 9A on July 9 2021 during the Return to Operations phase. The number of antennas in operation was 38. A few antennas among the 38 showed an elevated temperature during atmospheric calibration. The data were calibrated with the CASA \citep{CASA} pipeline version 2021.2.0.128 with CASA 6.2.1.7.  Two antennas were flagged out and $\sim$10\% of the data toward the phase calibrator were flagged. Despite a number of issues during calibration, the data were deemed science ready.

CIDA 9 was observed in Band 3 using six spectral windows as shown in Table 1. The main targeted lines are \ce{C^{18}O $J = 1-0$} (109.7822 GHz, $E_{\rm up}/k_{\rm B} = 5.27$ K, $\log_{10}(A_{\rm ul}) = -7.203$), \ce{^{13}CO $J = 1-0$} (110.2014 GHz, $E_{\rm up}/k_{\rm B} = 5.29$ K, $\log_{10}(A_{\rm ul}) = -7.198$), and \ce{CS $J=2-1$} (97.2710 GHz, $E_{\rm up}/k_{\rm B} = 7.05$ K, $\log_{10}(A_{\rm ul}) = -4.77$).  The properties of the molecular transitions are obtained from the CDMS \citep{cdms1,cdms2} and JPL \citep{jplline} catalogs.  The range of baselines was 122 m to 12.6 km. The data were finally imaged and self-calibrated using the latest CASA 6.4.  The phase center of the Band 3 data is ICRS 05:05:22.821 +25:31:30.426.  A single-phase self-calibration was performed using the aggregate pseudocontinuum.  The images were cleaned using \textsc{tclean} task in standard mode with Briggs \citep{briggs95} weighting of 0.5. The final images and their noise levels are tabulated in Table.~\ref{tab:imgs}.

The spectral lines were imaged after continuum subtraction via \textsc{uvcontsub}.  Due to the flux uncertainties of the phase calibrator and the high fraction of flagging, we adopt a 20\% flux uncertainty in Band 3 observations throughout this paper.  We have combined our data with published Band 6 observations \citep{flong18,manara19,rota22}.  The Band 6 data were corrected for the proper motion of CIDA 9A using the GAIA DR2 data \citep{gaiadr2, 
gaiadr2astrometry}.

\section{Results} \label{sec:results}

\begin{table}[]
    \centering
    \caption{Properties of the compact dust continuum emission at 93 and 233 GHz.  Flux densities ($S_{\nu}$) are calculated in the image plane by summing up the pixels with $I_{\nu} > 3\sigma$. }
    \label{tab:imgs}
    \begin{tabular}{c c c c c }
    \hline
    \hline
        Name & Beams & $rms$ & $S_{\nu}$ & $I_{\nu, peak}$  \\
        & \farcs, \farcs, \degr & mJy bm$^{-1}$ & Jy & mJy bm$^{-1}$ \\
    \hline 
    \multicolumn{5}{c}{Briggs 0.5} \\
    Band 3 & 0.14, 0.05, 3.7    & 0.02  & $1.8 \pm 0.4$ & 0.3\\
    Band 6 & 0.13, 0.10, 2.2    & 0.1   & $34 \pm 6$    & 2.8 \\
    \multicolumn{5}{c}{Briggs 0.5, Taper at 0\farcs{09}} \\
    Band 3 & 0.135, 0.135, 0    & 0.03  & $2.6 \pm 0.6$  & 0.4 \\
    Band 6 & 0.135, 0.135, 0    & 0.07  & $34 \pm 6$    & 3.5 \\
    \hline
    \end{tabular}
\end{table}

\subsection{Dust continuum emission around the primary} \label{sec:dustimg}

\begin{figure*}
    \centering
    \includegraphics[width=0.9\textwidth]{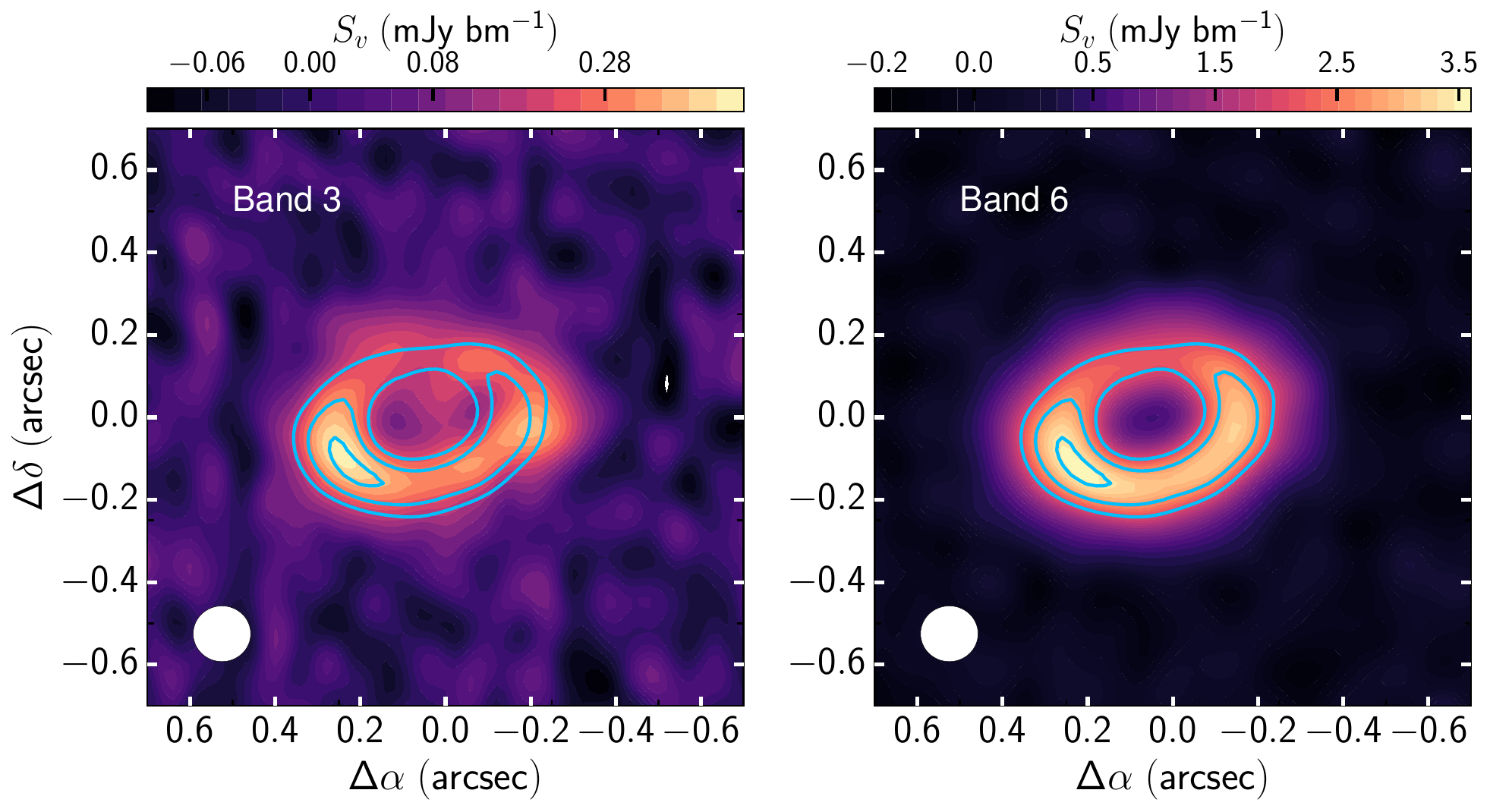}
    \caption{Dust continuum emission toward CIDA 9A in both Bands 3 ({\it Left}) and 6 ({\it Right}). The images are deconvolved using a tapering at 0\farcs{09} and are restored with a $0\farcs{135}\times0\farcs{135}$ beam.  Blue contour lines indicate the 30, 40, and 50 $\times \sigma$ levels of the Band 6 image to highlight the difference in the dust continuum peaks. The peak position in Band 3 is shifted clockwise with respect to the peaks that are seen in Band 6. The noise levels are tabulated in Table.~\ref{tab:imgs}.
    }
    \label{fig:dust}
\end{figure*}
\begin{figure}
    \centering
    \includegraphics[width=0.48\textwidth]{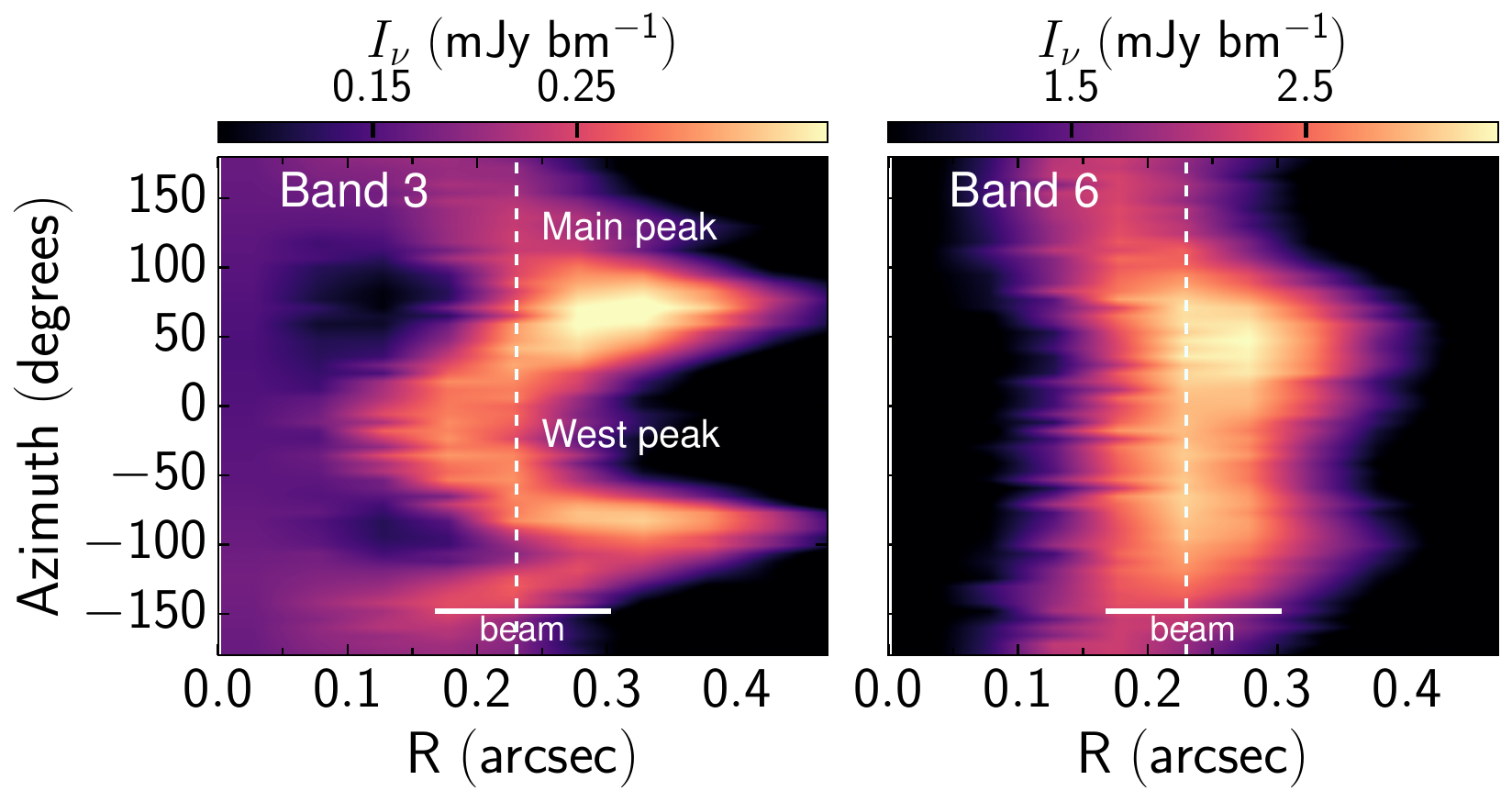} 
    \caption{Profile of the dust emission in both Bands 3 (left) and 6 (right). The vertical dash line  ($R = 0\farcs23$) shows the location of the emission peaks in Band 6 that corresponds to the dust ring location of \citet{flong18}.  The color scale is stretched differently with respect to Fig.~\ref{fig:dust} to highlight the dust emission peaks.  The coordinates have been deprojected using the inclination and position angle obtained in Section~\ref{sec:uvmodel}. 
    }
    \label{fig:dustprof}
\end{figure}

In order to compare the Band 3 and Band 6 images, the images are deconvolved including a $(u,v)$ tapering at 0\farcs{09} and a common 0\farcs{135} beam, to restore the final image. These tapered images are shown in Fig.~\ref{fig:dust}.  The original images at their respective native native resolution are shown in Appendix~\ref{app:A}. 

First, we analyze the data in the image plane to detect any radial and azimuthal 
shifts between Bands 3 and 6.  There are two dust concentrations: the main 
bright dust emission peak on the east side (hereafter, the main dust emission 
peak) and the fainter localized arc to the west (hereafter, the secondary peak). 
 The main dust emission peak in Band 3 seems aligned with the Band 6 image.  The 
deprojected dust emission profiles are plotted in Fig.~\ref{fig:dustprof} using 
the inclination and position angles found in Sect.~\ref{sec:uvmodel}.  
\citet{flong18} found that the ring is located at a deprojected distance of 
0\farcs{23} corresponding to 40 au (updated for the new Gaia DR3 distance).  The 
main peak seems to lie within the ring, without any noticeable radial shifts 
between the two observations. The secondary peak to the west seems to indicate a 
radial extension up to $\sim 0\farcs{2}$ with respect to Band 6 that is more 
than a single beam across.

An asymmetric dust continuum emission is observed at 96 GHz (3.1 mm), similar to 
that seen inthe Band 6 data \citep[1.3 mm,][]{flong18}. The emission at the 
east side of the disk is brighter by 16\%, with a significance level of 
2.3$\sigma$, than the west side of the disk.  The north and south variation is 
33\% in Band 3, which is $\sim 3 \sigma$, while it is 29\% in Band 6 ($\sim 13 
\sigma$).  In addition, the south side is brighter than the north side.  Our new 
Band 3 image shows more pronounced asymmetries in the east-west and north-south 
directions.  More of the dust emission in the ring is concentrated on the 
southeast side of CIDA 9A.  The companion CIDA 9B is not detected in Band 3 (see 
Appendix).  Here, we focus on the disk around the primary star CIDA 9A.

\begin{figure*}
    \centering
    \includegraphics[width=0.9\textwidth]{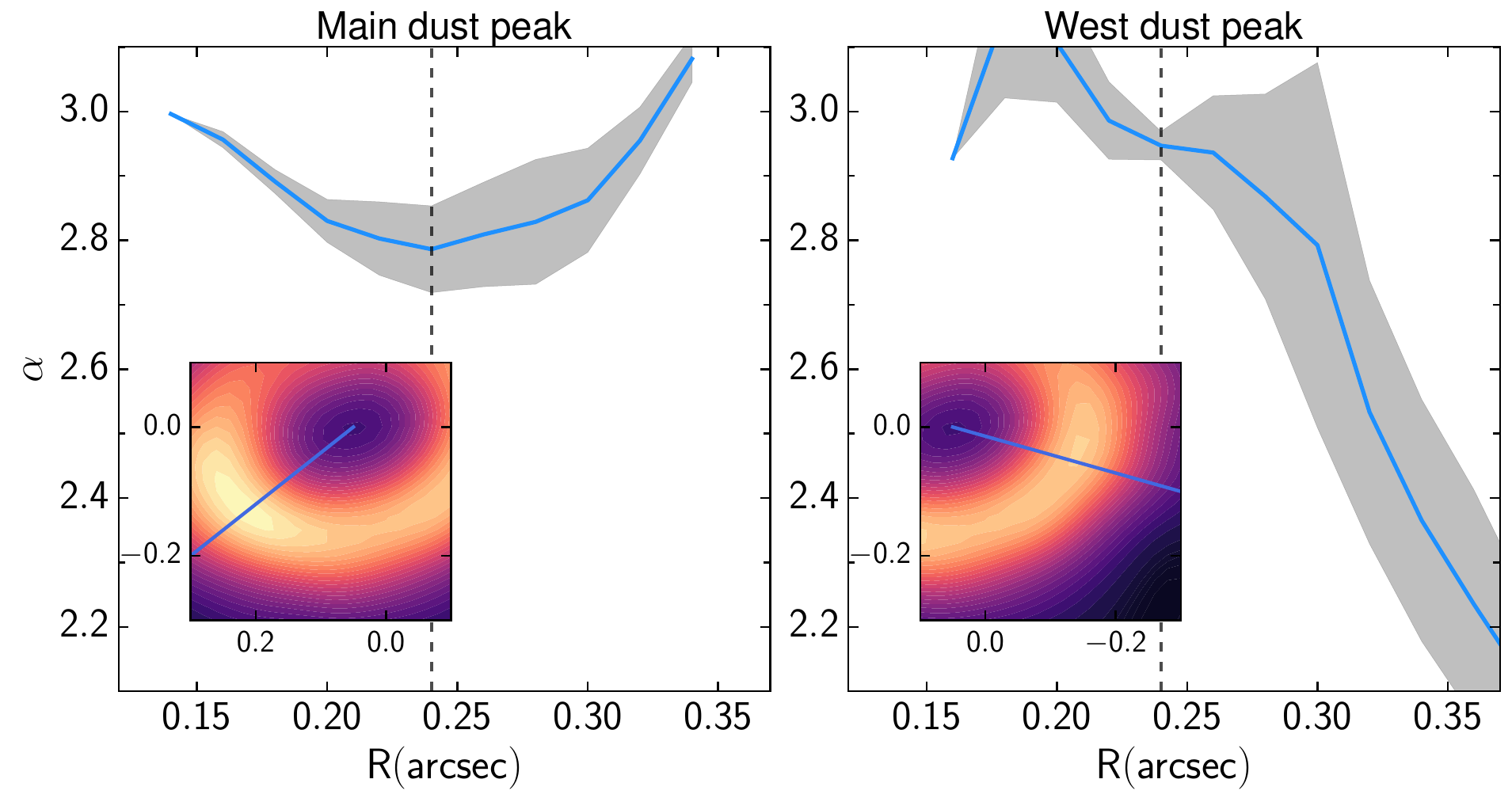}
    \caption{Dust spectral index profile of the dust emission peaks to the east 
(main, \textit{left}) and west (\textit{right}).  We plot the mean $\alpha$ values within a beam as a function of distance from the center in arcseconds in blue while the gray-shaded regions indicate the dispersion around the mean.  The vertical dash lines are placed at the rough locations of the peaks.  Insets show the zoom-in view to the emission peak that is associated with each panel along with the radial cut that is shown by the blue line.  The color scale in the insets is similar to Fig.~\ref{fig:dust}. 
    }
    \label{fig:dustspxprofile}
\end{figure*}

The dust grain distribution is usually estimated from the dust spectral index.  If the dust emission is optically thin, the multi-frequency dust spectral index $\alpha$ ($S_{\nu} \propto \nu^{\alpha}$) should be smaller than 3 to indicate the presence of larger dust grains \citep[with caveats as discussed by][]{testi14}.   To infer the dust grain population around the dust emission peaks, we need to understand the relationship between dust emission, the dust optical depth, and the dust spectral index \citep[e.g.,][]{cgonzalez19}.  In general, the dust emission is given by 
\begin{eqnarray}
    I_{\nu}  & = & B_{\nu} \left ( T_{\rm dust} \right ) \left [ 1 - \exp \left 
( -\tau  \right )\right ], 
\end{eqnarray}
where $\tau$ is the dust optical depth, and $B_{\nu}$ is the Planck function for a given dust temperature $T_{\rm dust}$.  The dust temperature profile is approximated as a power-law following the irradiated flared disk model of \citet{kenyon87} 
\begin{eqnarray}
    T_{\rm dust} (R) & = & T_{\star} \left ( \frac{R_{\star}}{R} \right )^{1/2} 
\phi_{\rm flaring}^{1/4}, 
\end{eqnarray}
where $\phi$ is the flaring angle of 0.05 \citep{dullemond04}, a stellar temperature $T_{\star}$ of 3592 K, and a stellar radius $R_{\star}$ of $\sim 1.2$ $R_{\odot}$ (values are obtained from \citealt{herczeg14} adjusted to a distance of 175 pc).  The dust optical depth depends on the surface mass density of the dust and the total dust extinction $\chi_{\nu} = \kappa_{\rm abs} + \sigma_{\rm sca}$. For simplicity, we consider dust emission from the dust absorption-dominated region.  In this limit, the dust emission is given by
\begin{eqnarray}
    I_{\nu} & = &  B_{\nu} \left ( T_{\rm dust} \right ) \left [ 1 - \exp \left 
( -\tau_0 \left( \nu / \nu_0 \right )^{\beta} \right )\right ]
\end{eqnarray}
with the dust absorption opacity slope $\beta$.  At the location of the ring, 
the dust temperature is $\approx 25$ K, and the dust spectral index $\alpha = 
\log \left ( I_{\nu_1} / I_{\nu_2} \right )/\log \left ( \nu_1/\nu_2 \right )$. 
Only in the case where $\tau_{\rm dust} < 1$, we can estimate the particle size 
distribution from $\alpha$, and the dust spectral index should be $\alpha 
\approx \beta + 2$.  As $\tau_{\rm dust}$ approaches 1, $\alpha$ approaches 2.  
In general, $\alpha$ is $\sim 3.7$ for interstellar medium dust grains and is 
$2\sim3$ if the dust grains have grown to larger sizes as $\alpha <$2 
could indicate non-thermal emission contribution.  The exact value depends on 
the composition and size distribution of the dust grains \citep{testi14}.

In order to estimate the dust optical depth, we compare the observed intensity with the expected dust emission such that $T_{\rm brightness} = T_{\rm dust} \left ( 1 - e^{-\tau_{\rm v}} \right )$ while considering that the dust emission is evenly distributed within the beam.  The average brightness temperatures around the two dust emission peaks are 3 to 5 K.  Using the expected dust temperature of 25 K, the optical depth in Band 3 $\tau_{\rm B3} \sim 0.1$ while $\tau_{\rm B6} \sim 0.3$ in Band 6 within a 0\farcs{135} beam.  The typical error in $\tau$ is $<0.1$, depending on the adopted temperature (15 -- 25 K) and 10\% error on the brightness temperature.  At the native resolution, the optical depth increases by a factor of two, indicating that the dust ring is still unresolved.  Therefore, the observed dust emission in Band 3 can be considered optically thin for this analysis while the dust emission in Band 6 is marginally optically thick, after considering the beam filling factor since the optical depth approaches 0.5 within a 0\farcs{135} beam.

We directly calculate the dust spectral index $\alpha_{\rm B6 - B3}$ in the image plane by evaluating $\log \left ( I_{B3} / I_{B6} \right )/\log \left ( \nu_{\rm B3}/\nu_{\rm B6} \right )$.  The value $\alpha_{\rm B6-B3}$ is evaluated using pixels with intensities $>5 \sigma$ in both Bands 3 and 6 images.  Figure~\ref{fig:dustspxprofile} shows the dust spectral index profile along the two dust emission peaks.  As a check, we also determined the spectral index from the concatenated observations.  To construct the spectral index, we first align Bands 3 and 6 data using \textsc{fixplanets}.  The concatenated visibilities are deconvolved using \textsc{tclean} with the Multi-term Multi-Frequency Synthesis (\textsc{mtmfs}) with three Taylor coefficients \citep{tsukagoshi22}.  These images and comparisons to the values evaluated in the image plane are shown in Appendix~\ref{app:A}.  For the purpose of this paper, we will present the spectral index values calculated from the images in  Fig.~\ref{fig:dust}.

As shown in Fig.~\ref{fig:dustspxprofile}, the localized lower spectral index of $\alpha_{\rm B6-B3} \sim 2.8$ is found near the main dust peak while the secondary peak shows an averaged $\alpha_{\rm B6-B3}$ of 2.9.  The standard flux uncertainty produced by the ALMA pipeline tends to be poorer than 10\% \citep{logan20b}.  However, our Band 3 observations were taken during the return to operations phase, which may lead to higher flux uncertainty as inferred from the large phase RMS.  The typical error of the spectral index is 0.2 (10\% flux uncertainty) and 0.3 (20\% flux uncertainty).  Despite the possible flux calibration issues, the major uncertainty lies in the parameters chosen during the image reconstruction using \emph{tclean}.  Depending on the parameters used during the deconvolution, the spectral index can vary by 0.5.  Nevertheless, a lower spectral index is found for the main dust emission peak with $\alpha < 3$.  Note that the images are deconvolved using a taper and a common beam.  The actual dust spectral index of the ring itself is a lower value if we use the images at their native resolution with higher noise levels.

\subsection{(u,v) modelling of the dust morphology around CIDA 9A} \label{sec:uvmodel}

\begin{figure*}
    \centering
    \includegraphics[width=0.98\textwidth]{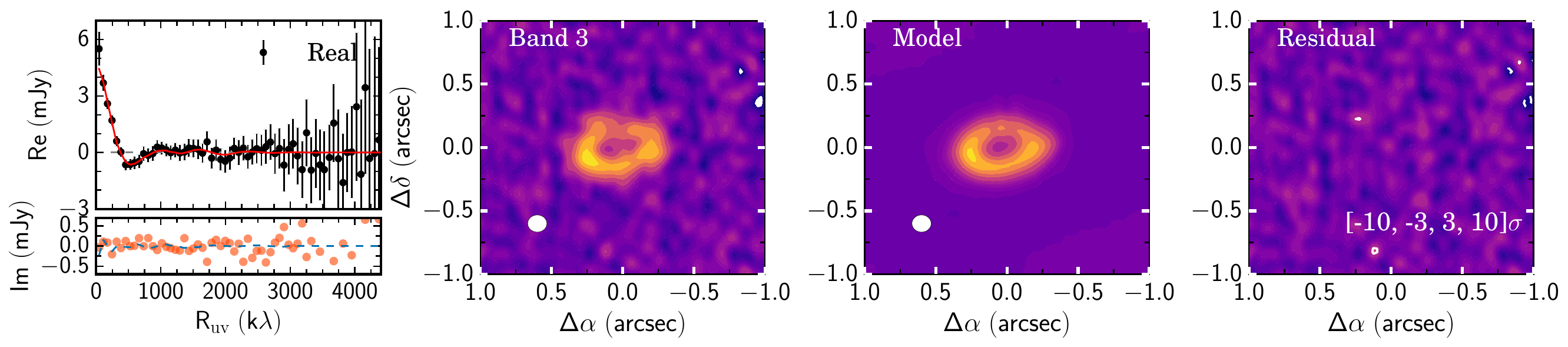} \\
    \includegraphics[width=0.98\textwidth]{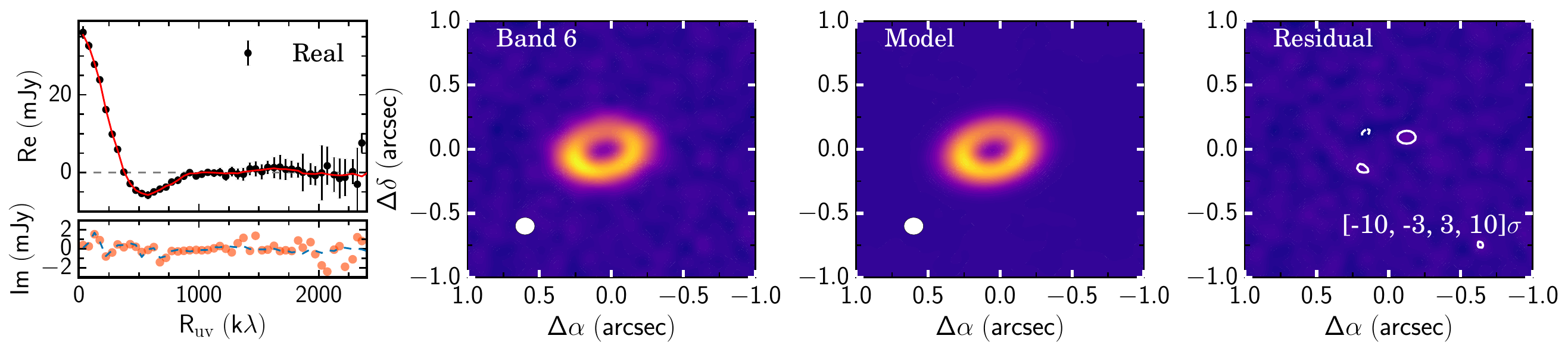} \\
    \caption{Best-fit symmetric Gaussian ring and two axisymmetric Gaussian ring model (GR2ARC) for both Bands 3 and 6. {\it Top:} The best-fit models for the Bands 3 data. {\it Bottom:} The best-fit model for the Bands 6 data. For each row, we show the binned visibilities, the original image, the model image, and the residual. For the residual image, the contours at -10, -3, 3, and 10 $\sigma$ are indicated by the white lines. 
    }
    \label{fig:dustModel}
\end{figure*}

\begin{table}
    \centering
    \caption{Location of the dust peaks from the $(u,v)$ fitting. }
    \label{tab:fits}
    \begin{tabular}{c c c c c c c }
    \hline
    \hline
        B & $R_{\rm 0}$ & $R_{\rm arc}$ & $\theta_1$ &  $\theta_2$ &
        $i$ & $ PA$ \\
            & au    & au & \degr &  \degr & \degr & \degr\\
    \hline 
    \multicolumn{5}{c}{Symmetric Gaussian Ring} \\
    3   &  42$^{+1}_{-1}$     & ...   & ... & ... & 48.6$^{+1}_{-1}$    & 
99.5$^{+1}_{-1}$  \\
    6   &  40$^{+2}_{-2}$     & ...   & ... & ... & 45.9$^{+0.2}_{-0.1}$ &  
104$^{+0.2}_{-0.2}$\\
    \multicolumn{5}{c}{Symmetric Gaussian Ring + 2 Arcs} \\
    3   &  42$^{+1}_{-1}$     & 40$^{+0.2}_{-0.2}$        & 36$^{+20}_{-16}$  & 
122$^{+20}_{-18}$  
        & 48$^{+1}_{-1}$     &  97$^{+2}_{-1}$\\
    6   &  42$^{+0.2}_{-0.2}$       & 37$^{+0.2}_{-0.2}$  & 58$^{+3}_{-2}$  
        & 146$^{+10}_{-19}$ & 45$^{+0.5}_{-1}$     &  110$^{+0.8}_{-4}$\\
    \hline
    \end{tabular}
\end{table}

\begin{figure}
    \centering
    \includegraphics[width=0.47\textwidth]{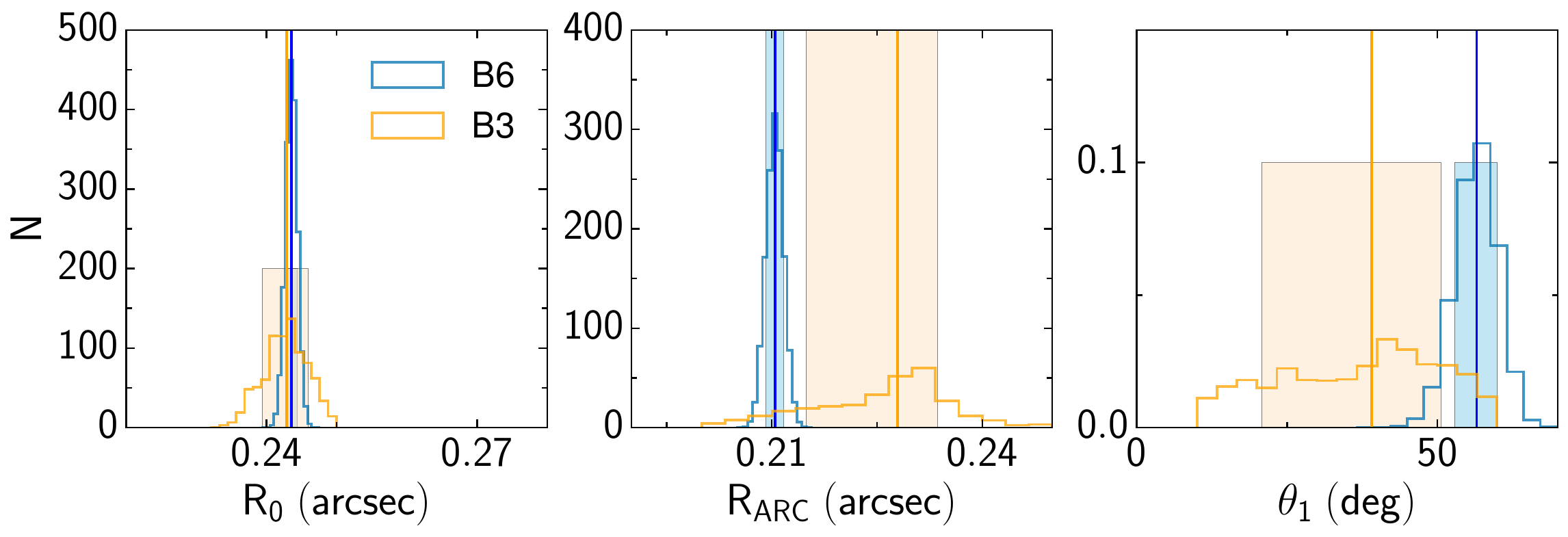} 
    \caption{Comparison between the best-fit values of the GR2ARC model in Bands 3 and 6.  The main parameters of the ring and the main arc are shown for simplicity. The mean and standard deviation for each of the parameters is plotted after taking into account the burn-in phase ($\sim 2000$) and an acceptance rate of 15\%. Band 6 results are shown in blue while Band 3 results are shown in orange. 
    }
    \label{fig:resultsGR2ARC}
\end{figure}

By fitting the $(u,v)$ data, we can accurately determine the location of the observed dust peaks in Bands 3 and 6 such that any azimuthal and radial shifts can be measured by comparing the best-fit parameters.  We fit the data with morphological models using \textit{Galario} \citep{galario}.  First, following \citet{flong18}, the data is fit with an axisymmetric Gaussian ring 
\begin{equation}
    I_{\rm Ring}(r) = A \exp \left [ - \frac{(R - R_{0})^2}{2 \sigma^2} \right ], 
\end{equation}
where the ring is centered at $R_0$ with a Gaussian width of $\sigma$. We confirm that the results of our fits to the Band 6 data are identical to \citet{flong18}, which are listed in Table.~\ref{tab:fits} (see App.~\ref{app:C}).

To recover any non-axisymmetric dust emission around CIDA 9A, we proceed to employ an asymmetric azimuthal Gaussian ring (hereafter, arc model) in addition to the axisymmetric Gaussian ring intensity model.  The arc model is based on a description of a vortex in a protoplanetary disk by \citet{lyra13} (see also \citealt{cazzoletti18}).  This particular model creates the arc feature $I_{\rm ARC} = I_{\rm 1} + I_{2}$, where
\begin{eqnarray}
    I_1 \left (r, \theta \right ) = A \exp \left [ - \frac{(R - R_{\rm arc})^2}{2 \sigma_r^2} \right ]\exp \left [ - \frac{(\theta - \theta_{1})^2}{2 
    \sigma_{\theta_1}^2} \right ]
\end{eqnarray}
and 
\begin{eqnarray}
    I_2 \left ( r, \theta \right )= A \exp \left [ - \frac{(R - R_{\rm arc})^2}{2 \sigma_r^2} \right ]\exp \left [ - \frac{(\theta - \theta_{2})^2}{2 
    \sigma_{\theta_2}^2} \right ], 
\end{eqnarray}
respectively, with $I_1$ describing the intensity profile clockwise with respect to the center of the arc $\theta_1$ ($\theta < \theta_1$, 0\degr is east) and $I_2$ is the intensity profile anticlockwise of $\theta_2$ located at the radius $R_{\rm arc}$.  The model creates arc-like features by separating the east and west sides of the Gaussian ring.  The free parameters are intensity $A$, radial position of the arc $R_{\rm arc}$, the width in radial direction $\sigma_r$, the angular position $\theta_1$ or $\theta_2$ where the arc corresponds to $I_1$ and $I_2$, the width in the position angle directions $\sigma_{\theta_1}$ and $\sigma_{\theta_2}$, position angle $PA$, inclination $i$, and the offset from the phase center ($\Delta \alpha$, $\Delta \delta$).  The total model is the symmetric Gaussian ring and two non-axisymmetric Gaussian ring models (arcs) such that $I_{\rm total} = I_{\rm Ring} +  I_{\rm ARC1} + I_{\rm ARC2}$ (hereafter GR2ARC).  For this paper, we have fixed the location of the arcs at a single $R_{\rm arc}$.  In total, we have 18 free parameters.

The best-fit parameters are determined using \textit{Galario} and the \textit{emcee} packages \citep{emcee} to explore the parameter space efficiently.  For each model, Gaussian and GR2ARC, we performed 10000 steps with a burn-in that is determined through the autocorrelation function ($\sim 1500-2000$ steps).  From these runs, the best-fit parameters are determined statistically using steps with an acceptance fraction of 0.2. We report the mean of the parameters, and their errors are calculated from the standard deviation from the mean.  We subtract the modeled visibilities from the observations using these best-fit values to create the residual maps.  The best-fit model and residual maps are shown in Figs.~\ref{fig:dustModel} and \ref{fig:dustModelapp} for the GR2ARC and Gaussian model, respectively.  The best-fit values are listed in Table~\ref{tab:fits}. The models capture most of the observed features with 1--3 $\sigma$ residuals as seen from the last panels in Figs.~\ref{fig:dustModel} and \ref{fig:dustModelapp}. These residuals are minimized by the GR2ARC model as shown in Fig.~\ref{fig:dustModel}.

The axisymmetric Gaussian ring model indicates that the ring in Band 3 is radially shifted by $0\farcs{01}$ or 2 au with an error $\sim 2$ au with respect to the Band 6 visibilities. The center of the observations is determined with uncertainties of $\sim 0\farcs{01}$. Considering these uncertainties, the rings in Bands 3 and 6 are located at the same radius based on the axisymmetric Gaussian models.  We will then focus on the best-fit parameters of the GR2ARC model.  Table~\ref{tab:fits} shows the best-fit parameters with restricted locations for the axisymmetric Gaussian ring and arcs.  If we let the radial location $R$ to be free, the center of the ring ($R_{0}$) is located at 44${\pm 1}$ au for Band 6 while it is 44$^{+{7}}_{-6}$ au in Band 3.  The main dust peak is fitted with the arc located at 37$\pm 0.2$ au in Band 6 while it is at 40$\pm 0.6$ au in Band 3.  Similarly, the center position of the two observations is fitted with statistical uncertainties of $\sim 1-2$ au.  Considering the uncertainties in the center position and proper motion, we report an upper limit to the radial shift $\lesssim 3$ au between Bands 3 and 6.

\begin{figure}[!]
    \centering
    \includegraphics[width=0.48\textwidth]{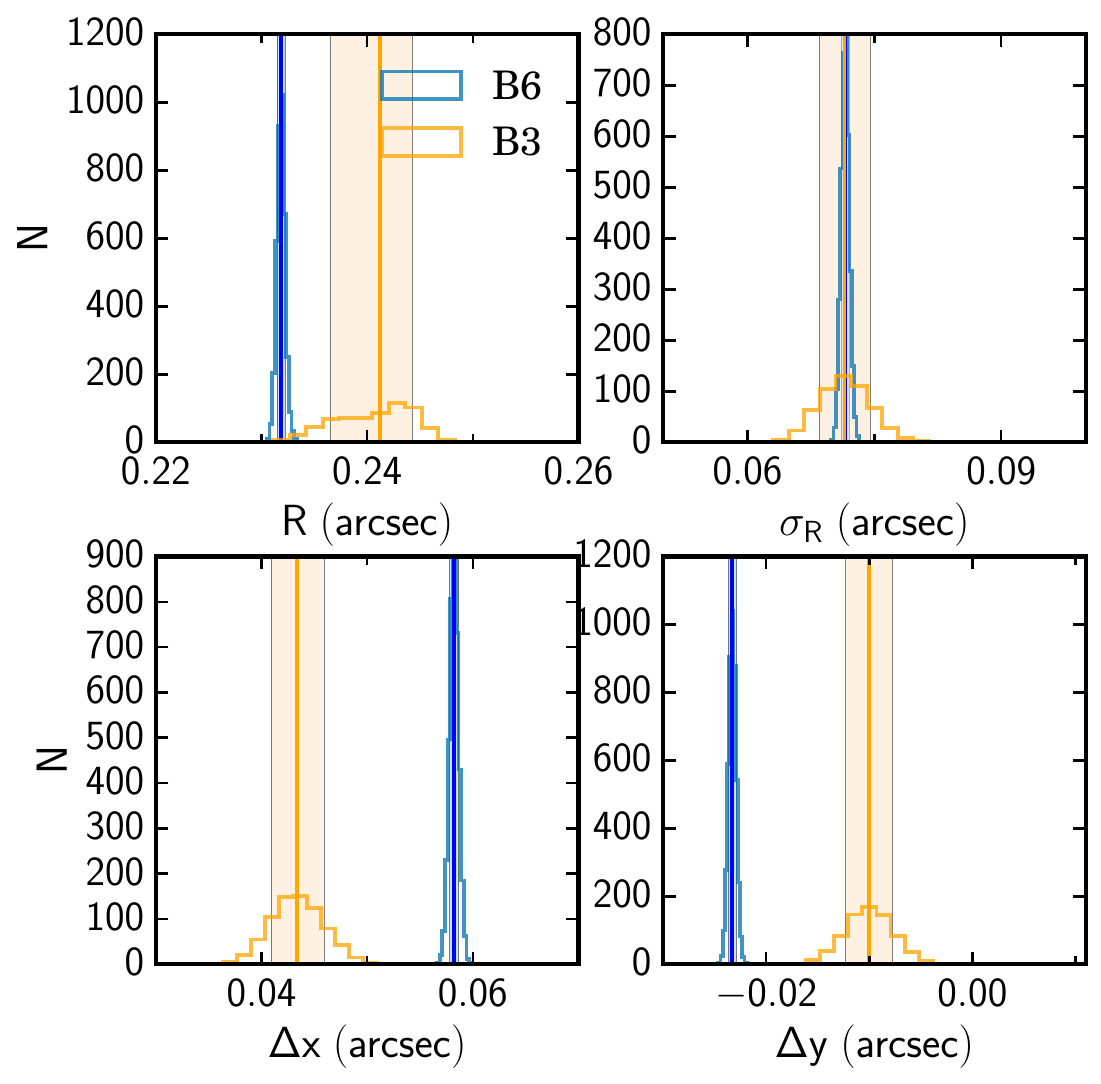} 
    \caption{Similar to Fig.~\ref{fig:resultsGR2ARC} but for the Gaussian Ring models.
    }
    \label{fig:resultsGR}
\end{figure}

The main dust peak shows an azimuthal shift between Bands 3 and 6. The peak in 
Band 3 is located at 36$^{+20}_{-16} \ ^{\circ}$ while the Band 6 data is 
centered at 58$^{+3}_{-2} \ ^{\circ}$, which results in 22$^{+5}_{-36}$\degr\ 
shift.  We fit the observed visibilities twice using the GR2ARC model.  The 
first run allows all of the parameters, including the position of the arc 
$R_{\rm arc}$ to be free, which implies that $I_{1}$ can be radially shifted 
with respect to $I_{2}$. The second set of runs restricts the arcs, $I_{1}$ and 
$I_{2}$, to be at the same radius $R_{\rm arc}$.  In comparison, without 
restricting the radial locations of the ring and the two arcs (the first set of 
runs), the angular difference is $\sim 80\degr$.  Meanwhile, the location of the 
second arc that describes the west peak is similar in both Bands 3 and 6.  With 
these $(u,v)$ modeling, we find a tentative azimuthal shift in the main dust 
peak between Bands 3 and 6 since the shift is seen in Fig.~\ref{fig:dustprof} 
and recovered through the visibilities analysis with non-negligible 
uncertainties. We calculated the Bayesian information criterion (BIC) for each 
of the intensity models, which is given by $n_{\rm par} \log N_{\rm data} - 2 
\log \mathcal{L_{\rm max}}$ where $n_{\rm par}$ is the number of parameters, 
$N_{\rm data}$ is the number of data points, and $\mathcal{L_{\rm max}}$ is the 
maximum likelihood function of the model given the best-fit parameters.  Even 
though the GR2ARC model returns a lower residual map in Band 6 (see 
Fig.~\ref{fig:dustModel}), its BIC value is much greater than the Gaussian ring 
model by a factor of 3.  This implies that we cannot conclusively 
determine that GR2ARC is the better model for the observations presented here.

Using these models, we constructed the model images within \emph{CASA} to determine the spectral index.  The dust spectral index $\alpha$ obtained from the model images is $\sim 2.6-2.9$ at the dust continuum peaks.  These values are consistent with the observed dust spectral index in Fig.~\ref{fig:dustspxprofile}. Using the models, we can further estimate the dust spectral index of the dust ring itself without deconvolution.  The underlying intensity models indicate that the spectral index at the main dust peak is $\sim 2$.  The optical depth of the best-fit images shows that $\tau_{\rm B6} = 2$ while $\tau_{\rm B3} \sim 0.4$.  Therefore, the models indicate that the main dust peak is optically thick with a low spectral index.

\subsection{Molecular line emission}

\begin{figure*}
    \centering
    \includegraphics[width=0.98\textwidth]{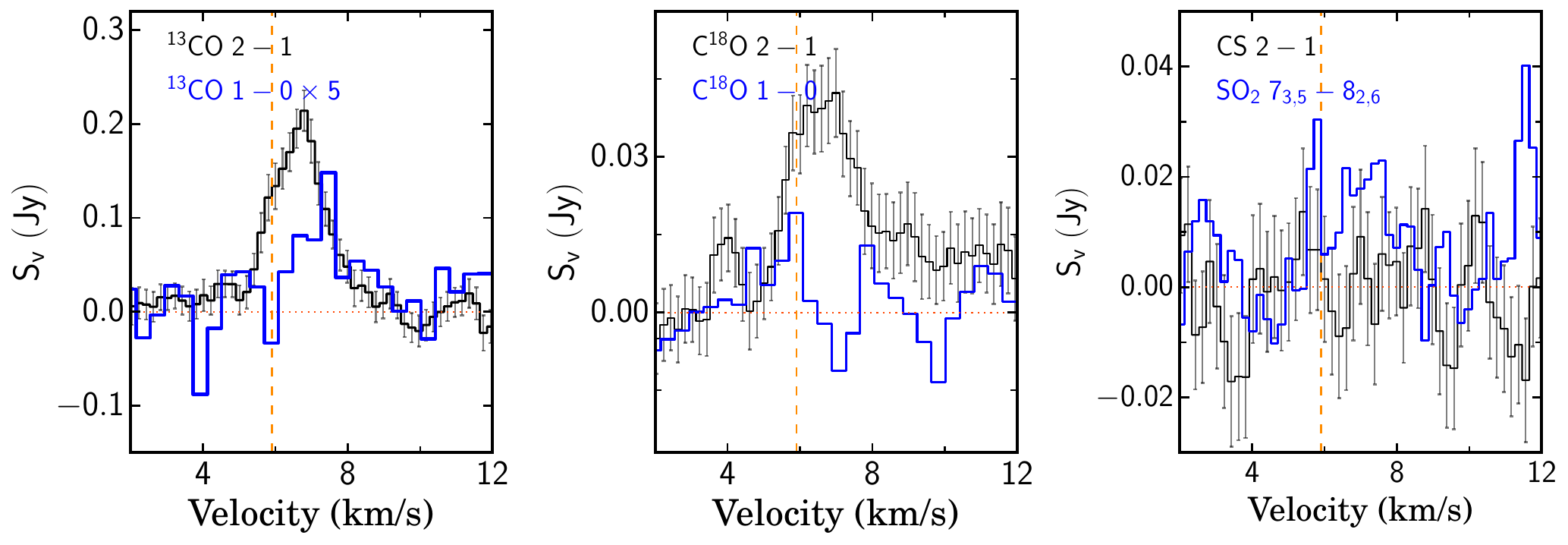} \\
    \caption{Molecular line profiles around CIDA 9A. As a reference, the \ce{^{13}CO} and \ce{C^{18}O} $J = 2-1$ are shown in black lines in the left and middle panels. Their $J = 1-0$ counterparts are shown in blue. In addition, CS 2--1 and \ce{SO2} $7_{3,5} - 8_{2,6}$ are plotted in the right panel. These spectral lines are extracted with \textit{gofish} adopting the stellar and disk parameters of \citet{rota22}. 
    }
    \label{fig:lineprofiles}
\end{figure*}

Along with the dust continuum, the Band 3 observations also targeted \ce{^{13}CO} $J=1-0$ (110.2014 GHz), \ce{C^{18}O} $J=1-0$ (109.7822 GHz), and \ce{CS} $J=2-1$ (97.2710 GHz).  We have also included the \ce{^{13}CO} $J=2-1$ and \ce{C^{18}O} $J=2-1$ observations from \citet{manara19} and \citet{rota22} for completeness.  We have used a Keplerian masking using the parameters of \citet{rota22} ($i = 56^{\circ}$, PA = $284.2^{\circ}$) to extract the molecular lines in Bands 3 and 6 using \emph{GoFish} \citep{Teague_gofish2019}. Figure~\ref{fig:lineprofiles} shows the integrated line emission from the primary disk.  We clearly detect the weak \ce{^{13}CO} 1--0 and \ce{C^{18}O} 1--0 emissions using the Keplerian mask.  The low-$J$ lines suffer from foreground absorption.   Meanwhile, both \ce{SO_2} and CS line emissions are tentatively detected.

Since the velocity-based masking is able to extract the disk emission, we proceed to fit the $J=2-1$ line with {\it eddy} \citep{eddy}.  For our fitting, we fix the inclination to $i = 47^{\circ} \pm 2$ based on the orientation of the dust disk in Section~\ref{sec:uvmodel}.  The dust disk and gas disk as observed from the rotational transitions in the sub-mm should not be strongly misaligned hence we adopt the inclination of the dust disk.  The first moment maps are obtained using {\it bettermoments} \citep{bettermoments} with the quadratic method including a 3$\sigma$ clip.  The first moment maps are then fitted using {\it eddy} with 2000 steps including a 10\% burn-in.  We obtain a stellar mass of 0.62 $M_{\odot}$ using the \ce{^{12}CO} 2--1 line and 0.72 $M_{\odot}$ with the \ce{^{13}CO} 2--1 line.  The typical error on the stellar mass is $<0.1 \ M_{\odot}$. If we adopted an inclination of $56^{\circ}$ instead \citep{rota22}, emission models would provide a better fit to the high spatial resolution CO lines, as obtained along with the data presented in \citet{flong18}, but a worse fit with larger residuals in the short baseline data presented here. With the longer baseline configuration, a significant part of the line emission at 3--5 km s$^{-1}$ is not recovered.  For the purpose of this paper, we adopt the lower inclination fit solution with a stellar mass between 0.62 -- 0.72 $M_{\odot}$, inclination of $47^{\circ} \pm 2$, and a position angle between $277^{\circ} - 283^{\circ}$. 

\begin{table}[]
    \centering
    \caption{Comparison of the integrated line flux densities in Jy km s$^{-1}$.  \citet{rota22} reported the total fluxes obtained from their cumulative flux technique.  We report the total integrated flux density inside of 1\farcs{5} and 0\farcs{5} circles using \emph{goFish} adopting a 20\% error. }    
\begin{tabular}{c| c  c  c  c}
    \hline
    Source  &  \ce{^{13}CO} 1--0    & \ce{^{13}CO} 2--1     & \ce{C^{18}O} 1--0     & \ce{C^{18}O} 2--1 \\
    \hline
    Rota+22 & ...      & 0.49$^{+0.02}_{-0.02}$   & ...   & 0.12$^{+0.01}_{-0.01}$   \\
    1\farcs{5}  & $<0.08$  & 0.51$\pm 0.14$    & $<0.1$      &  $0.32 \pm 0.15$      \\
    0\farcs{5}  & $0.030 \pm 0.01$   & $0.23 \pm 0.03$       &  $0.04 \pm 0.02$        &  $0.10 \pm 0.01$     \\
    \hline
    \end{tabular}

    \label{tab:integratedline}
\end{table}

Using these parameters, we integrate the $J=1-0$ and $J=2-1$ lines of \ce{^{13}CO} and \ce{C^{18}O}.  The molecular line emission is detected between 2 and 10 km s$^{-1}$.  The integrated line flux density is calculated by integrating the line profile obtained with \emph{goFish} with the maximum radius of 1\farcs{7} as obtained by \citet{rota22}.  The velocity range for the integration is estimated from the line profiles as shown in Fig.~\ref{fig:lineprofiles}.  For the weak lines CS and \ce{SO2}, the line emission is integrated from 4 to 10 km s$^{-1}$, where the tentative \ce{SO2} line is seen.  For the \ce{^{13}CO} $J=2-1$ lines, the total integrated flux density is derived using the images from \citet{rota22}.  The error is given by the number of channels and the $rms$ such that $1.2 \times \sqrt{N_{\rm chan}} \times rms$.  The first factor of 1.2 is taking into account the 20\% flux uncertainty.  Note that \ce{^{13}CO} $J=2-1$ line presented in \citet{rota22} are obtained with shorter baselines configuration than the data presented in \citet{flong18}.

By adopting the higher stellar mass, the integrated line flux density and the peak intensity of \ce{^{13}CO} 1--0 and SO$_2$ are higher by more than 20\%.  These data support a stellar mass for CIDA 9A of at least 0.6$M_{\odot}$.  Unfortunately, the data is not deep enough to search for any kinematical disturbances within the millimeter cavity.  Due to the low S/N of the gas lines observations, we do not detect any broad molecular line emission that can be associated with molecular winds. The moment maps of these lines can be found in App.~\ref{app:B}.

\section{Discussion} \label{sec:discuss}

Our high-spatial resolution observations of the transitional disk around CIDA 9A in Band 3 show similar asymmetric substructures as the previous Band 6 data.  In order to bring out the dust substructures, we have deconvolved the images with a 0\farcs{09} taper and a 0\farcs{135} circular beam.  The aim is to find differences in the observed substructures between Bands 3 and 6 data.  Our data indicate a speculative azimuthal shift in the location of the main dust emission peak between Bands 3 and 6 and an upper limit of 3 au radial shift characterized by a localized spectral index $\alpha < 3$.  We will focus on discussing the main mechanisms that can form the transition disk around CIDA 9A.

A wide variety of mechanisms can be invoked to explain the large asymmetric disk 
with a cavity in the sub-mm around CIDA 9A. As shown in \citet{flong18}, CIDA 9A 
does not indicate a dust cavity through its spectral energy distribution.  The 
major difference between CIDA 9A and other sources with known large asymmetric 
dust traps is the properties of the central star. Most of the previous 
detections of dust asymmetries are around Herbig and/or Sun-like stars 
stars, while there are only a few detections of an asymmetric dust ring around 
$< 1 \ M_{\odot}$ stars\footnote{The adopted mass for this paper relies upon 
pre-main sequence evolutionary tracks that produce higher masses.  Many previous 
measurements of other disk-hosting stars rely on models that yield lower masses 
for the same parameters.} \citep{gonzalezruilova20, hashimoto21}.

The millimeter cavity in CIDA 9A can be carved by either an embedded planet, disk wind, and/or dead zone. To favor between a vortex generated by an embedded planet and a dead zone, we are looking for potential radial and azimuthal differences between Bands 3 and 6 observations, although this does not provide a definitive answer about the origin of the cavity. In the case of a massive protoplanet (the mass depends on the disk viscosity and scale height), we expected a drop of gas and dust inside the millimeter cavity \citep[e.g.,][]{lubow06,zzhu11,rosotti13, duffell15,kanagawa17,villenave19} and an azimuthal shift between small and dust grains due to the vortex \citep{Baruteau2016}. If disk wind is the main driver of the cavity formation \citep{alexander07, rosotti13, ercolano17, garate21}, we expect a large gas-depleted cavity, and the presence of broad molecular line emission as the gas is entrained from the millimeter cavity. Lastly, a pressure trap generated at the edge of the dead zone can form a cavity and a millimeter dust ring \citep{flock15}. \citet{pinilla2019} showed that there exists a radial shift between small and large dust grains due to the pressure trap generated by a dead zone.

We step through several possible scenarios of how the observed millimeter could have been formed and discuss whether their observable signatures are observed.

\paragraph{Giant protoplanet:} Although most of the observed dust asymmetry is explained by proto-planets embedded in the disk or the gap, this explanation is not straightforward for CIDA 9A.  Millimeter dust can be confined in a vortex near the dust gap or cavity formed by a Jupiter-type planet \citep{rdong15b}.  The observed total emission from the disk around CIDA 9A is 30--40 mJy at 233 GHz, which translates to 0.18--0.25 $M_{\rm Jup}$ of solids adopting a dust temperature of 20 K and a dust grain absorption opacity of 10 cm$^{2}$ g$^{-1}$ at 1000 GHz with a frequency dependence $\beta=1$ \citep{beckwith90}.  As mentioned earlier, the dust temperature at the dust ring's location is $\sim 25$ K, which lowers the total dust mass.  A giant planet could have been formed when the disk was more massive in the first 1 Myr year \citep[e.g.,][ and references therein]{drazkowskappvii}.

\cite{rota22} showed that gas is still present in the dust cavity of CIDA 9A.  On the other hand, our high-spatial CO data observations do not have the spatial resolution and sensitivity to infer the underlying gas distribution for comparison with the dust distribution.  Recently, \citet{pinilla21} and \citet{hashimoto21} showed cavities that indicate the possible presence of giant planets around very low-mass stars.  In comparison to CIDA 9A, the ZZ Tau IRS disk still has about 0.07--0.15 $M_{\rm jup}$ (0.18--0.25 $M_{\rm jup}$ for CIDA 9A) dust mass around it.  With the revised higher stellar mass obtained for CIDA 9A, it is still possible that there is a protoplanet inside the millimeter cavity since the disk mass could be much higher in its earlier stages.

If a planet was present, in addition to the difference in gas and dust distributions, a gas vortex could form at the outer edge of the cavity \citep{dvborro07, ataiee13, pinilla22}.  At the location of the ring of CIDA 9A, the orbital time is $\sim 300$ years. \citet{pinilla22} found that the dust concentration inside such a  vortex is still present after 3000 orbits, or $9 \times10^{5}$ years.  Since CIDA 9 is likely 3--10 Myr old, it seems that the dust concentration should be maintained by other mechanisms.  By considering realistic planet formation time-scales, a double-peaked millimeter emission that is observed toward CIDA 9A can be obtained by a single vortex after long times of evolution \citep{hammer2019}. The observed azimuthal shift is still consistent with this scenario.

The final observable signature of an embedded protoplanet is the spiral arms in the scattered light.  Unfortunately, CIDA 9A is faint in 2MASS\footnote{The faint near-IR photometry with 2MASS may be the consequence of a faint epoch or confusion between the primary and secondary.  The source is probably brighter in most epochs.} and difficult to be followed up by ground-based IR instruments. Therefore, we could not obtain any scattered light images.  At the moment, the suitable way to rule out the presence of a giant protoplanet is to obtain high-spatial resolution and high-sensitivity CO isotopologues (\ce{^{12}CO}, \ce{^{13}CO}, and \ce{C^{18}O}) observations toward the millimeter cavity around CIDA 9A to measure the underlying gas distribution with respect to the dust.

\paragraph{Winds.}  From the detection of CO gas emission inside the millimeter 
cavity, it places limits on how photoevaporation may form the millimeter cavity 
in CIDA 9A. On the other hand, our molecular line data are not sufficient to 
constrain the presence of any fast wind.  It is not possible to constrain 
whether a wind is still actively driving the evolution of the millimeter 
cavity toward CIDA 9A without additional gas line (atomic and molecular) 
observations to constrain the wind.

\paragraph{Dead zone.}  
Our results indicate a tentative azimuthal shift between the smaller dust grains 
as traced in Band 6 ($\sim 230$ GHz, 1.3 mm) and larger dust grains in Band 3 
($\sim 90$ GHz, 3.1 mm).  If the azimuthal shift is $\sim 20\degr$ as shown by 
the $(u,v)$ model, it is in agreement with the hydrodynamical simulations of gas 
and dust with a vortex in \cite{Baruteau2016}, where larger grains (here traced 
with Band 3) are expected to be shifted ahead of the vortex in the azimuthal 
direction. If this is the case for CIDA 9A, this implies that the disk rotation 
is clockwise. As shown in \cite{Baruteau2016}, such a shift can also produce a 
double-peaked azimuthal emission.  The models from \cite{Baruteau2016} assume an 
initial perturbation in the gas density profile that triggers the vortex 
formation and that can have different origins, such as planets, 
photoevaporation, and dead zones. The current data does not seem to favor any of 
these origins. The dead zone models presented by \cite{pinilla2019} predict a 
radial shift of the millimeter emission at different wavelengths (peak closer at 
longer) in the that is opposite to the prediction in the planet-disk interaction 
models. In CIDA 9A,  there is an indication of a very small radial shift, but it 
is too small to give any firm conclusion on the cavity origin based on these 
observations.

In order to investigate the processes that shape the disk around CIDA 9A, much 
deeper ALMA observations in multiple frequency bands are required.  In addition, 
deeper molecular gas emission that targets the molecular wind and warm gas 
within the millimeter cavity are needed to constrain the role of 
photoevaporative wind in shaping the disk evolution.  In particular, high 
spatial resolution Band 9 observation would help to completely rule out the 
presence of a vortex at the edge of the dead zone by constraining the radial 
shift between Bands 3, 6, and 9.  The millimeter dust ring produced at the edge 
of a dead zone is predicted to show an optically thick emission from the small 
dust grains that are radially shifted with respect to the longer wavelength 
emissions. Furthermore, a more detailed hydrodynamical evolution of multiple 
mechanisms is needed to compare with these new observations of disks. 

\paragraph{Spectral index of CIDA 9A}

The spectral index can be used as an observational diagnostic of dust growth 
when the millimeter emission is optically thin. In the case of CIDA 9A, a lower 
value of the spectral index in the radial direction is found along the main dust 
peak. The low spectral index between the two bands can also indicate a higher 
optical depth.  On the other hand, the observed spectral index $\alpha> 2$.  The 
dust spectral index around the main dust peak is close to 3 outside of the main 
dust peak, and the value gradually decreases toward the center of the main dust 
peak.  This suggests that particles have grown to millimeter-sizes at this 
location, as expected from a vortex {\citep{birnstiel2013}}. The spectral index 
increases with distance from the main dust peak as expected from dust evolution 
models,  and as it has been observed in several protoplanetary disks 
\citep[e.g.,][]{perez2012,tazzari2016,long2020}, and in disks with clear 
asymmetric rings \citep{casassus2015, vandermarel2015}.

For the western dust peak, the spectral index shows a slight decrease at this 
radial position, but it continues decreasing radially, suggesting that grains 
are large in the outer parts of the disks.  Unlike its brighter counterpart, 
this secondary peak seems to be inconsistent with a  vortex, because otherwise, 
the large grains would have concentrated in the vortex and only small grains 
would remain outwards (increasing the spectral index) as observed for the main 
dust peak.  Multi-wavelength observations of CIDA 9A at higher sensitivity and 
resolution are needed to confirm this unusual behavior of the spectral index.

\section{Summary and conclusions}

Multi-band observations of protoplanetary disks have been essential in 
constraining the physical mechanisms responsible for forming transition disks.  
The end stages of disks and how they disperse are crucial in understanding both 
disk evolution and planet formation.  This paper presents Bands 3 and 6 ALMA 
observations of the transition disk around an M-dwarf CIDA 9A.  The main results 
and conclusions can be found below:
\begin{itemize}
    \item We detected an asymmetric disk that can be modeled as a Gaussian ring 
and two arcs in our Band 3 data.  With the higher spatial resolution, the disk 
in Band 3 looks as asymmetric as in Band 6.  The secondary disk CIDA 9B is not 
robustly detected in Band 3.  The dust emission consists of the main dust 
concentration to the east or the main dust peak and the secondary dust peak to 
the west.

    \item With the two frequencies, we find a localized dust spectral index that 
dips to $\alpha_{\rm B3-B6} \sim 2.8$ at the main dust emission peak. The dust 
spectral index is calculated using the dust continuum images.  The dust emission 
seems to be optically thin in Band 3, while it is marginally optically thick 
$\tau \gtrsim 0.3$ in Band 6.  With these results, we cannot conclusively show 
that the large dust grains are concentrated at the dust emission peaks.

    \item Both Bands 3 and 6 data have been analyzed in the $(u,v)$ space by 
fitting the visibilities with an axisymmetric Gaussian ring and a combination 
of a Gaussian ring and two Gaussian arcs model (GR2ARC).  These fits indicate 
that the dust emission peaks are not radially shifted with respect to each 
other.  However, the models indicate a tentative azimuthal shift of 
$\sim22^{+5}_{-36}$\degr\  between the peak position in Bands 3 and 6 that 
requires further investigation.

    \item Our data also contain CO molecular lines, \ce{SO2}, and CS.  We can 
recover the \ce{^{13}CO}  and \ce{C^{18}O} $J=1-0$ lines using velocity-based 
masking.  On the other hand, CS $J=$ 2--1 and \ce{SO2} lines are tentatively 
detected within the millimeter cavity.  With these molecular data, we constrain 
the stellar mass to be around 0.62--0.72 $M_{\odot}$ adopting an inclination of 
$i\sim 47\pm2 \degr$ as derived from the $(u,v)$ model of the dust continuum 
data.

    \item The tentative azimuthal shift of  $\sim22^{+5}_{-36}$\degr\  is 
consistent with a dust trap caused by a vortex.  On the other hand, we cannot 
determine the cause of the vortex in the disk around CIDA 9A.  Deeper 
observations in both dust continuum and molecular lines are needed to 
investigate whether the vortex is generated by an unseen 
planet or other hydrodynamical instabilities.  
\end{itemize}

This paper makes use of the following ALMA data: ADS/JAO.ALMA \#2019.1.01270.S, 
ADS/JAO.ALMA \#2016.1.01164.S, and ADS/JAO.ALMA \#2018.1.00771.S. ALMA is a 
partnership of ESO (representing its member states), NSF (USA) and NINS (Japan), 
together with NRC (Canada), MOST and ASIAA (Taiwan), and KASI (Republic of 
Korea), in cooperation with the Republic of Chile. The Joint ALMA Observatory is 
operated by ESO, AUI/NRAO and NAOJ. 
DH is supported by Center for Informatics and Computation in Astronomy (CICA) grant and grant number 110J0353I9 from the Ministry of Education of Taiwan. DH also acknowledges support from the National Science and Technology Council of Taiwan through grant number 111B3005191.
Support for F.L. was provided by NASA through the NASA Hubble Fellowship grant \#HST-HF2-51512.001-A awarded by the Space Telescope Science Institute, which is operated by the Association of Universities for Research in Astronomy, Incorporated, under NASA contract NAS5-26555.
G.R. acknowledges support from the Netherlands Organisation for Scientific Research (NWO, program number 016.Veni.192.233) and from an STFC Ernest Rutherford Fellowship (grant number ST/T003855/1).  
GJH and YFS are supported by the National Natural Science Foundation of China grant 12173003. D.J.\ is supported by NRC Canada and by an NSERC Discovery Grant.
CFM is funded by the European Union under the European Union’s Horizon Europe Research \& Innovation Programme 101039452 (WANDA). FM\'e has received funding from the European Research Council (ERC) under the European Union's Horizon 2020 research and innovation program (grant agreement No. 101053020, project Dust2Planets). Views and opinions expressed are however those of the author(s) only and do not necessarily reflect those of the European Union or the European Research Council. Neither the European Union nor the granting authority can be held responsible for them.
This project has received funding from the European Union’s Horizon 2020 research and innovation programme under the Marie Skłodowska-Curie grant agreement No 823823 (DUSTBUSTERS).

\vspace{5mm}
\facilities{ALMA}

\software{astropy \citep{astropy18},  Galario \citep{galario}, Gofish \citep{Teague_gofish2019},  emcee \citep{emcee}, Splatalogue \citep{splatalogue}}

\appendix

\section{Continuum images} \label{app:A}

\begin{figure*}
    \centering
    \begin{tabular}{cc}
    \includegraphics[width=0.98\textwidth]{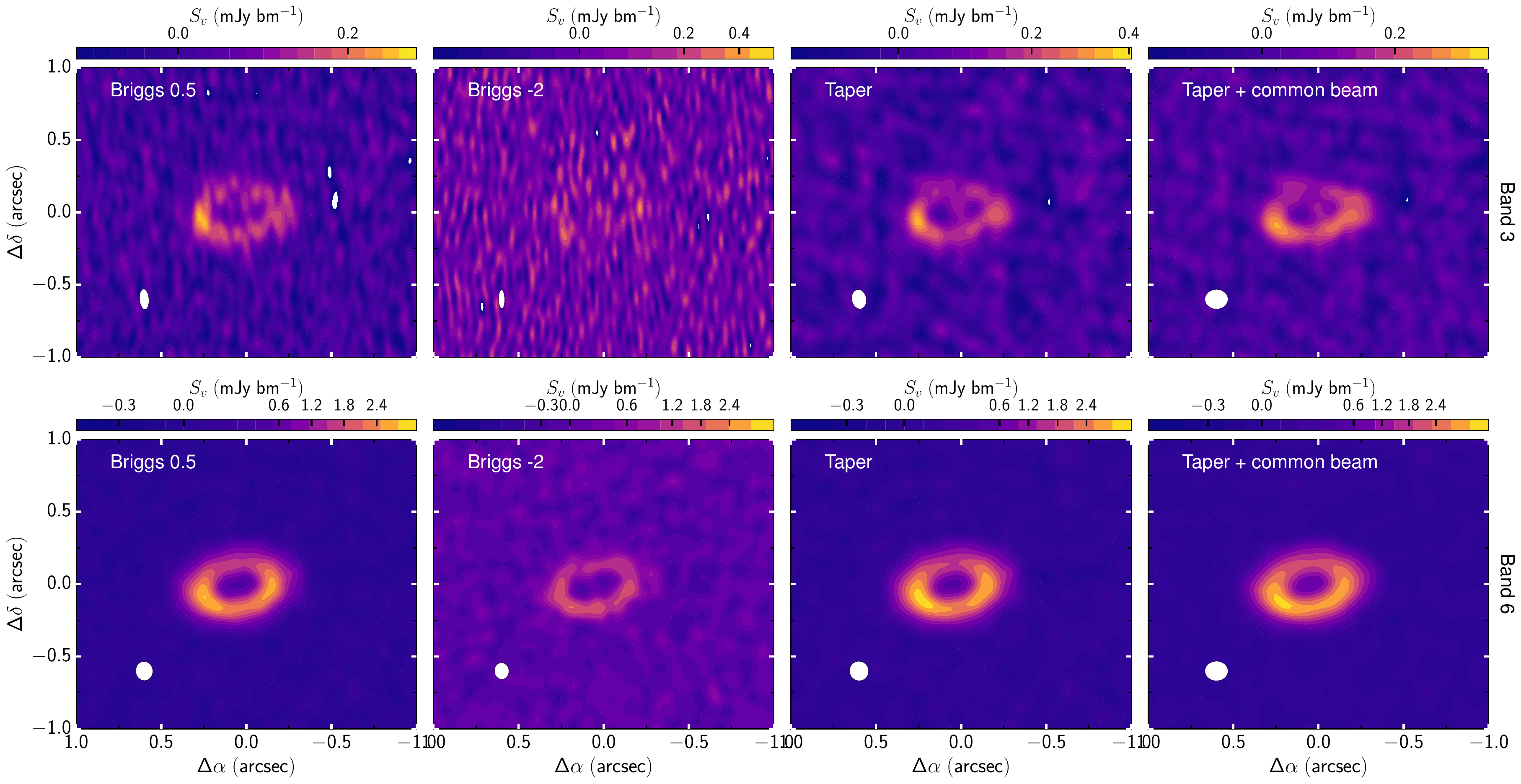}
    \end{tabular}
    \caption{
    Dust continuum images of CIDA 9A. {\it Top: } Band 3 images. {\it Bottom: 
} Band 6 images.  These images are cleaned with {\it Briggs} weighting.  The 
tapered images (taper) are constructed with {\it Briggs} weighting of 0.5 and a 
tapering at 0\farcs{08}.  The panels that are indicated by 'Taper + common beam' 
are images that are restored with a common beam of 0\farcs{135}. 
    }
    \label{fig:cida9aimages}
\end{figure*}

\begin{figure*}
    \centering
    \begin{tabular}{cc}
    \includegraphics[width=0.98\textwidth]{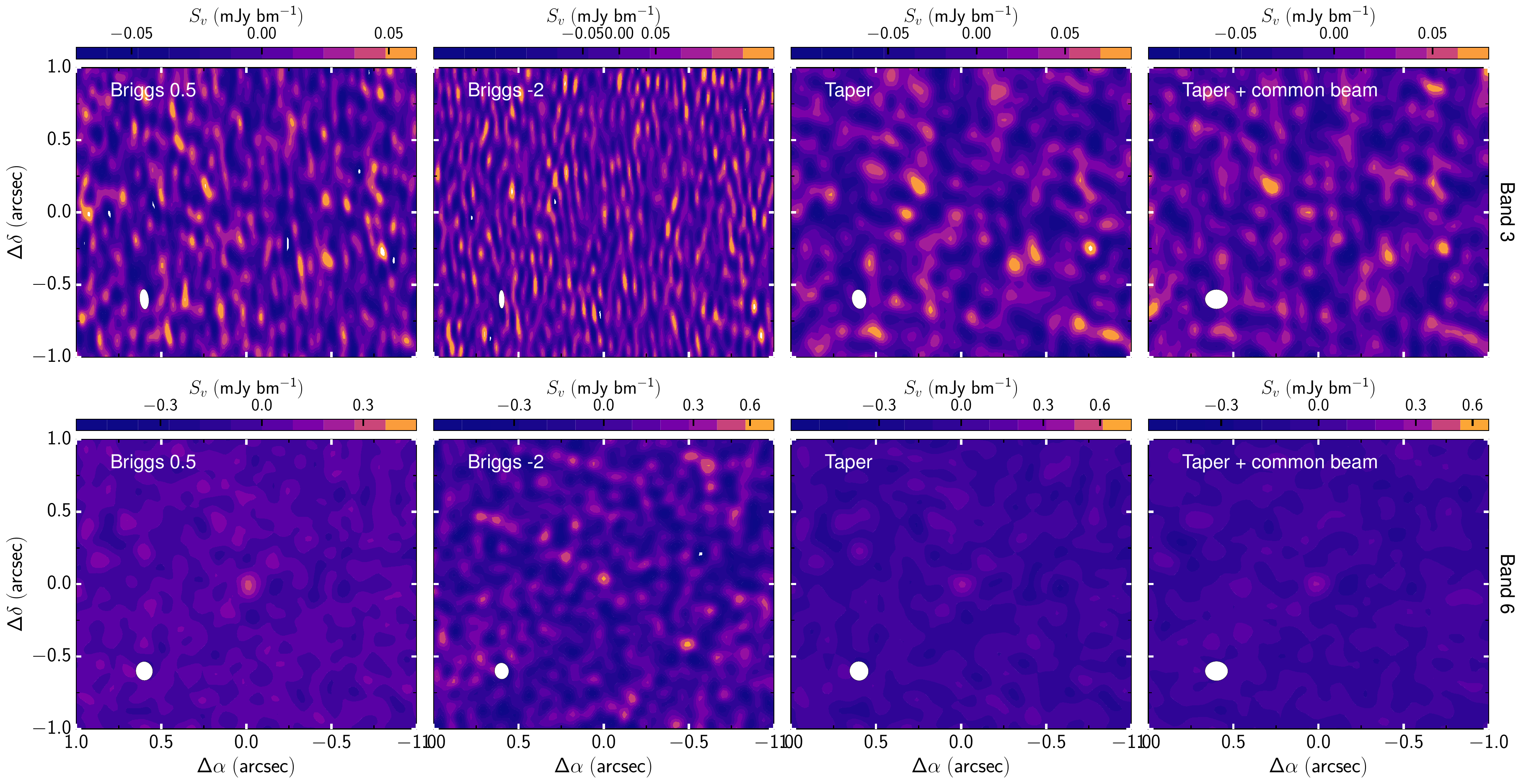}
    \end{tabular}
    \caption{
    Similar to Fig.~\ref{fig:cida9aimages} for CIDA 9B. 
    }
    \label{fig:cida9bimages}
\end{figure*}

We have imaged both Bands 3 and 6 data separately with {\it CASA} \emph{tclean}.  In order to analyze the two images in a consistent manner, we presented the results above based on  the images that are deconvolved with a 0\farcs{09} taper and a common restoring beam.  Fig.~\ref{fig:cida9aimages} shows CIDA 9A as imaged using various deconvolution parameters including \emph{Briggs} weighting and the fixed restoring beam.  The structures are similar within the different images.

The secondary star CIDA 9B is located 2\farcs{35} away to the direction of the north east of the primary disk \citep{manara19}.  The dust continuum emission in Band 6 from the disk around CIDA 9B is barely detected at a $\sim 3 \sigma$ level with our imaging.  \citet{manara19} reported a detection at 0.32 mJy in Band 6 which is at 5$\sigma$ level in a $\sim 0\farcs{14}\times 0\farcs{11}$ beam.  The secondary disk is shown in Fig.~\ref{fig:cida9bimages} centered on the coordinates reported by \citet{manara19}.   As shown, CIDA 9B is detected in Band 6 (1.3 mm) but it is not clearly detected in Band 3 (3.1 mm).  The upper limit in Band 3 is $31$ $\mu$Jy within a 0\farcs{135} beam and $23$ $\mu$Jy as determined from the image with Briggs weighting of 0.5.  The upper limit to the dust emission in Band 3 is 0.20 mJy within 0\farcs{1}.  With these flux densities, the lower limit to the spectral index of CIDA 9B is $\sim 3.4$ with an error of 0.4 that takes into account a 20\% flux error.

\begin{figure*}
    \centering
    \begin{tabular}{cc}
    \includegraphics[width=0.98\textwidth]{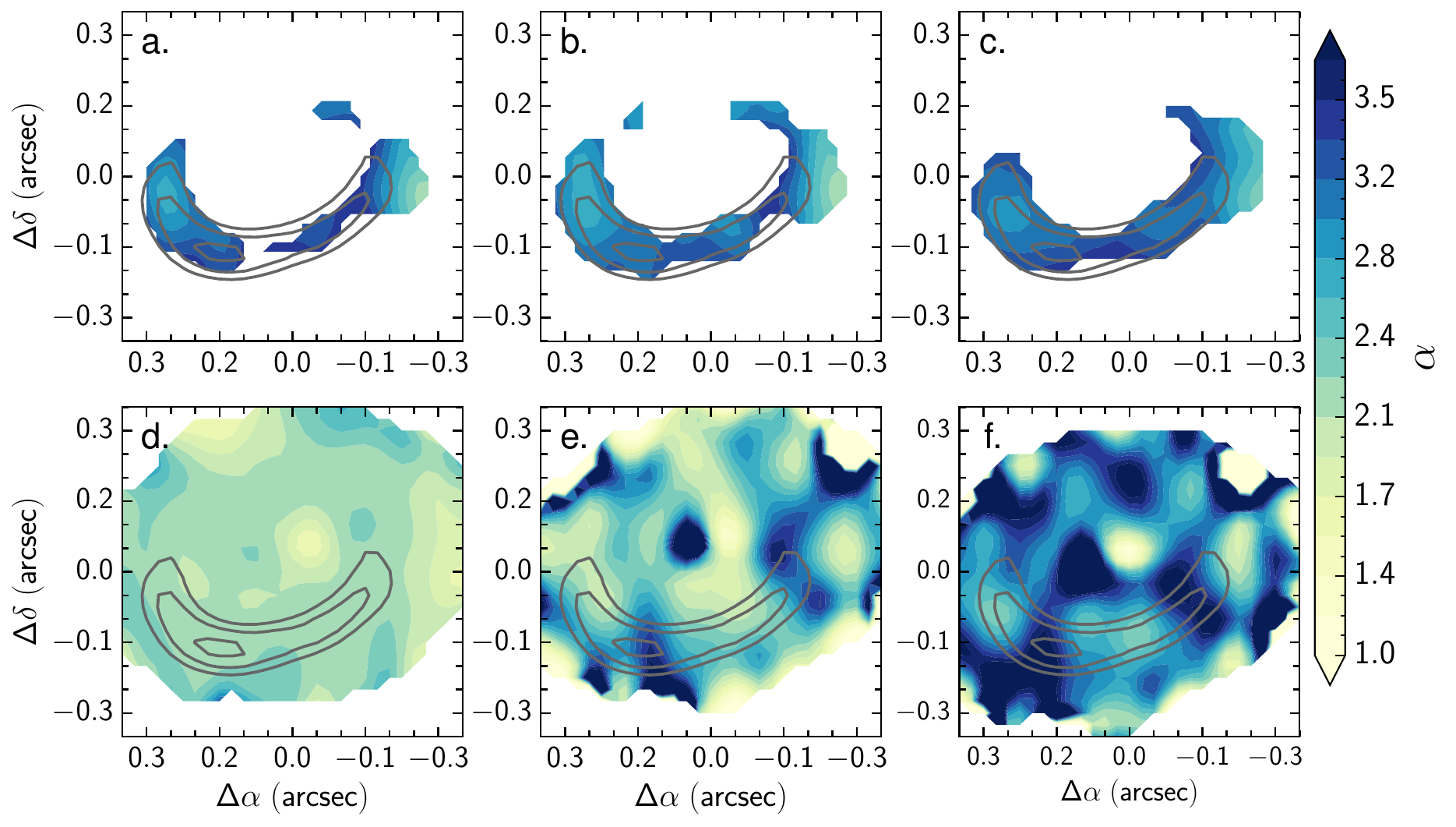}
    \end{tabular}
    \caption{
    Spectral index maps of the primary disk. The top row (panels a, b, and c) shows the spectral index $\alpha$ as derived in the image plane. Panel a is the spectral index map of the dust continuum images constructed with a tapering at 0\farcs{09} with \textit{mfs} and Hogbom deconvolver.  Panel b shows the map using \textit{mtmfs} and a taper while panel c shows the map in the case of the combination of tapering and a common beam of 0\farcs{135}. The second row shows the spectral index map as produced by \textit{tclean} \textit{mtmfs}. Panels d, e, and f shows the $\alpha$ map with increasing number of Taylor coefficients from 2 to 4. The gray contours indicate the dust continuum emission of the concatenated Bands 3 and 6 visibilities at 35, 40, and 45 $\sigma$. 
    }
    \label{fig:cida9aalpha}
\end{figure*}

In the main part of this manuscript, we evaluated the dust spectral index profile using the images in Fig~\ref{fig:dust}.  For completeness, Fig.~\ref{fig:cida9aalpha} shows the complete dust spectral index maps as evaluated using a different set of deconvolution parameters including \emph{mtmfs}.  We use $\alpha = \frac{\Delta I}{\Delta \nu}$ to calculate the spectral index.  Panels a, b, and c show the $\alpha$ map using the values that are derived from the images.   The differences between panels a, b, and c are due to the different parameters that are adopted during deconvolution with \emph{tclean}.  Panel a shows the result from deconvolution using \emph{mfs} with a tapering at 0\farcs{09} and \emph{Hogbom} deconvolution. Panels b and c use the \textit{mtmfs} deconvolver while keeping the taper at 0\farcs{09}. The image in panel c is restored with a common beam of 0\farcs{135}.

It is clear that the spectral index map is highly dependent on the adopted deconvolution options.  The images that are produced with \textit{mtmfs} using two Taylor terms tend to produce $\alpha \sim 2$ which is indicative of optically thick emission.  This value is in strong contrast to $\alpha \sim 2.8$ which is the value obtained by deriving the spectral index from the images.  The difference is very crucial in interpreting the grain size distribution around the dust continuum emission. An $\alpha \sim 2$ indicates an optically thick emission while $\alpha \sim 2.8$ is a moderate grain growth.

The major differences could be due to the fact that we are attempting to derive a spectral index using two frequencies.  However, ALMA should provide high-quality data that can be used to derive the spectral index using the different spectral windows in Bands 3 and 6.  Following \citet{tsukagoshi22}, we tested the spectral index that is derived with \textit{mtmfs} using a higher number of Taylor coefficients.  We find that the indices that are obtained using three and four Taylor coefficients are more consistent with those values obtained in the image plane.

\section{Moment maps} \label{app:B}

\begin{figure*}
    \centering
    \includegraphics[width=0.98\textwidth]{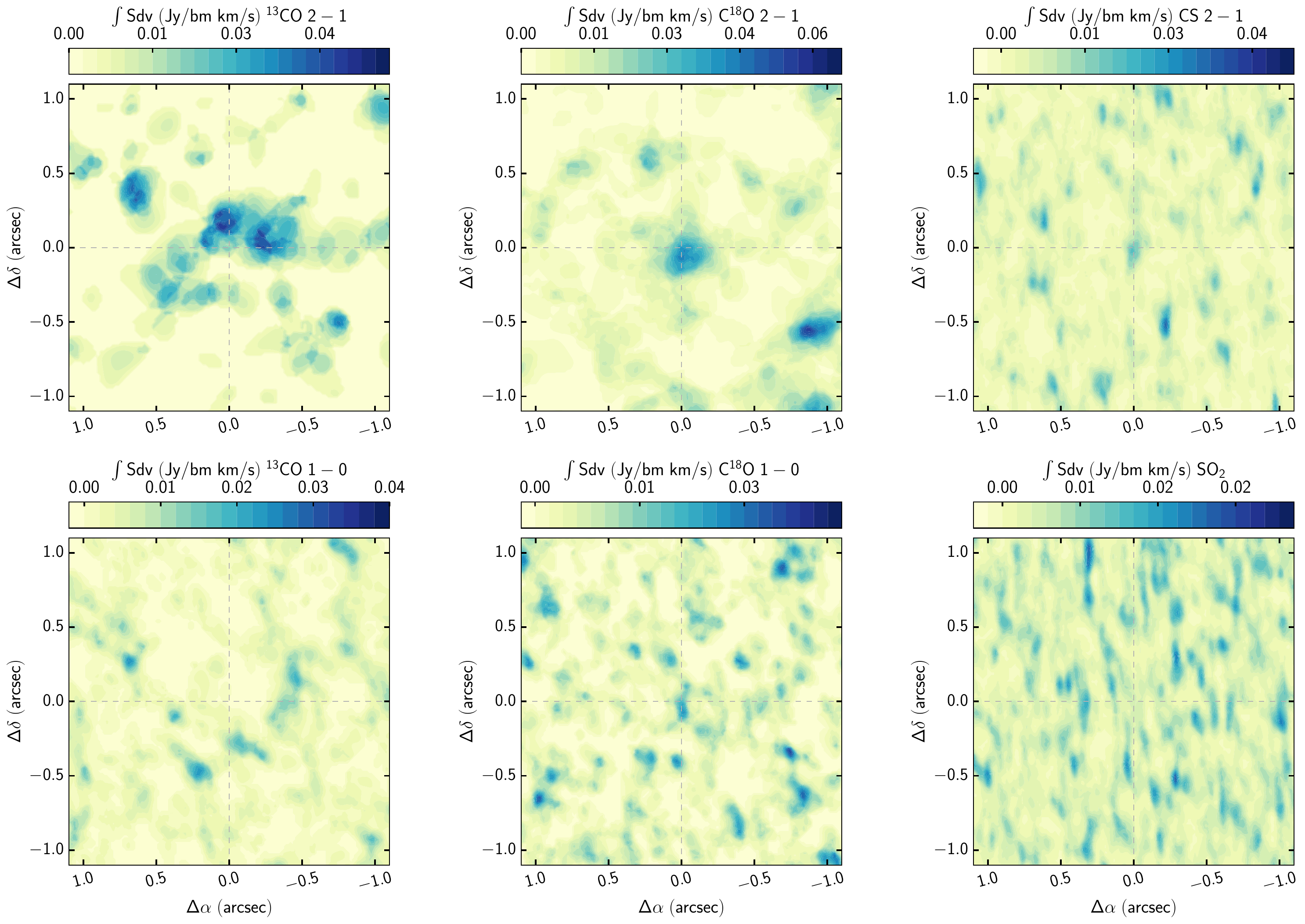} \\
    \caption{Integrated line maps (zeroth moment) of \ce{^{13}CO} $J=$2--1, \ce{C^{18}O} $J=$2--1, CS $J=$2--1, \ce{^{13}CO} $J=$1--0, \ce{C^{18}O} $J=$1--0, and \ce{SO2} $7_{3,5} - 8_{2,6}$. The spectral cubes are integrated from 2 to 10 km/s. Only emissions above 2$\sigma$ is considered. 
    }
    \label{fig:mommaps}
\end{figure*}

\begin{figure*}
    \centering
    \includegraphics[width=0.98\textwidth]{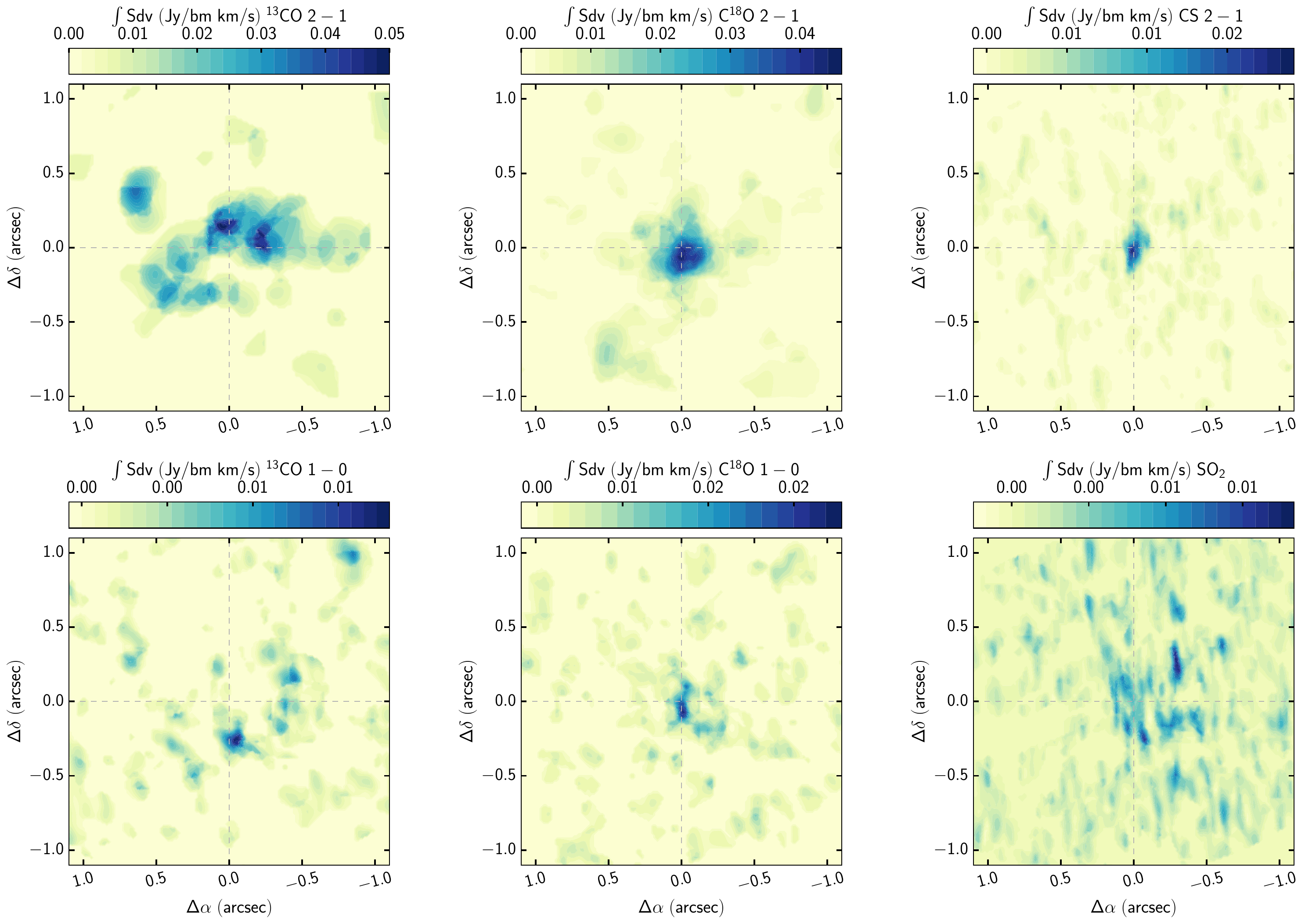} \\
    \caption{Similar to Fig.~\ref{fig:mommaps} with the addition of a velocity-based masking. We adopted the $i = 47\degr$ and a stellar mass of $0.7 M_{\sun}$.
    }
    \label{fig:mommapsKep}
\end{figure*}

Due to the low signal-to-noise data on the targeted gas lines in Band 3, we did not include the moment maps in the main section of the paper. Moment maps of the Band 6 data have been published in \citet{rota22}. Figure~\ref{fig:mommaps} shows the zeroth moment maps of the major species while fig.~\ref{fig:mommapsKep} show the moment maps constructed with the Keplerian masking. As explained in the results section, we adopted an inclination that was derived by fitting the dust continuum emission.

\section{Results from the ($u,v$) modelling} \label{app:C}

\begin{figure*}
    \centering
    \includegraphics[width=0.98\textwidth]{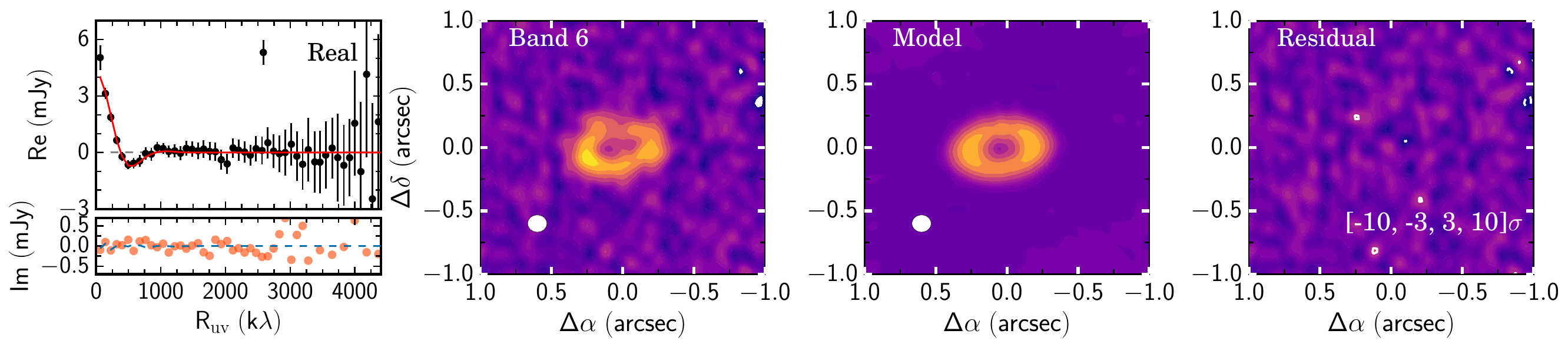} \\
    \includegraphics[width=0.98\textwidth]{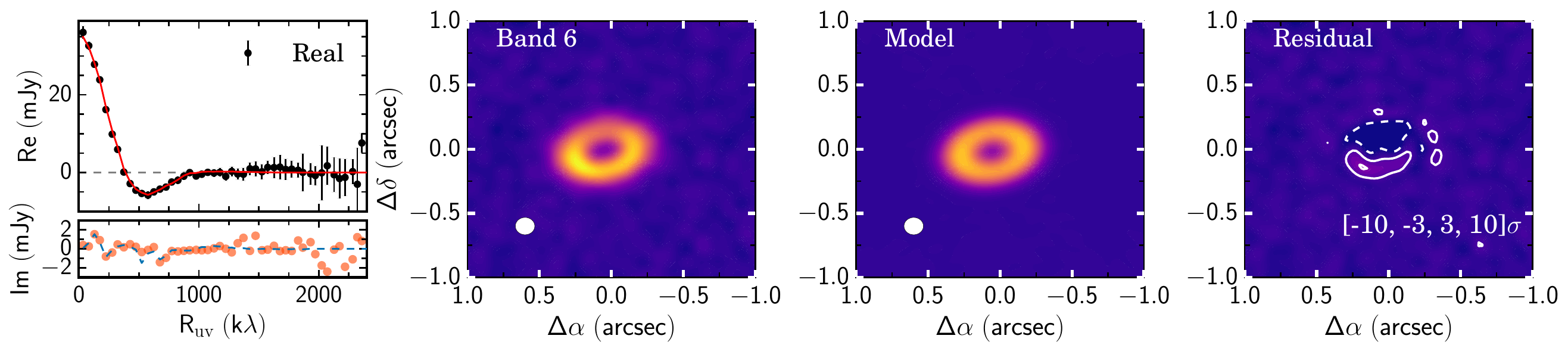} \\
    \caption{Best-fit models of the dust morphology in Bands 3 and 6. {\it Top:} The best-fit models for the Bands 3 data. {\it Bottom:} The best-fit model for the Bands 6 data. For each row, we show the binned visibilities, the original image, the model image, and the residual. For the residual image, the contours at -10, -3, 3, and 10 $\sigma$ are indicated by the white lines. 
    }
    \label{fig:dustModelapp}
\end{figure*}

We have presented the results of the fits using the GR2ARC model in the main text. 
In particular, we have also used the axisymmetric Gaussian ring model to fit the 
observed substructures to be consistent with the results presented by 
\citet{flong18}.  Figure~\ref{fig:dustModel} shows the axisymmetric Gaussian 
ring model.

\bibliographystyle{aasjournal}
\bibliography{CIDA9B3.bib}

\end{document}